\documentclass[aps,prl,amsmath,amssymb,floatfix,twocolumn,amsmath,superscriptaddress,twocolumn,nofootinbib,tighten,letterpaper]{revtex4-2}
\usepackage[colorlinks,linkcolor=blue,citecolor=blue,urlcolor=blue]{hyperref}
\usepackage{multirow}
\usepackage{subfigure}
\usepackage{color}
\usepackage{mathrsfs}
\usepackage{hyperref}
\usepackage[normalem]{ulem}
\usepackage{bm}

\usepackage[utf8]{inputenc}
\usepackage{amssymb}
\usepackage{amsmath}
\renewcommand\vec[1]{\ensuremath\boldsymbol{#1}}

\usepackage{amsfonts, relsize, color}
\usepackage{graphics}
\usepackage{graphicx}
\usepackage{subfigure}
\usepackage{color}
\usepackage{comment}

\begin{document}
\title{Quantized electrical, thermal, and spin transports of non-Hermitian clean and dirty two-dimensional topological insulators and superconductors}

\author{Sanjib Kumar Das}
\affiliation{Department of Physics, Lehigh University, Bethlehem, Pennsylvania, 18015, USA}

\author{Bitan Roy}
\affiliation{Department of Physics, Lehigh University, Bethlehem, Pennsylvania, 18015, USA}

\date{\today}

\begin{abstract}
From lattice-regularized models, devoid of any non-Hermitian (NH) skin effects, here we compute the electrical ($\sigma_{xy}$), thermal ($\kappa_{xy}$), and spin ($\sigma^{sp}_{xy}$) Hall, and the electrical ($G_{xx}$) and thermal ($G^{th}_{xx}$) longitudinal conductivities for appropriate NH planar topological insulators and superconductors related to all five non-trivial Altland-Zirbauer symmetry classes in their Hermitian limits. These models feature real eigenvalues over an extended NH parameter regime, only where the associated topological invariants remain quantized. In this regime, the NH quantum anomalous and spin Hall insulators show quantized $\sigma_{xy}$ and $G_{xx}$, respectively, the NH $p+ip$ ($p \pm ip$) pairing shows half-quantized $\kappa_{xy}$ ($G^{th}_{xx}$), while the NH $d+id$ pairing shows quantized $\kappa_{xy}$ and $\sigma^{sp}_{xy}$ in the clean and weak disorder (due to random pointlike charge impurities) regimes. We compute these quantities in experimentally realizable suitable six-terminal setups using the Kwant software package. But, in the strong disorder regime, all these topological responses vanish and with the increasing non-Hermiticity in the system this generic phenomenon occurs at weaker disorder.       
\end{abstract}

\maketitle

\emph{Introduction}.~Topological crystals displaying the hallmark bulk-boundary correspondence in terms of robust gapless modes living on their boundaries~\cite{Hasan2010, Qi2011, Chiu2016, kanemele2006, BHZ2006, FuKaneMele2007, Fukane2007, moorebalents2007, rahulroy2009, Ryu2010, Kitaev2009, Schnyder2008, Fu2011, Slager2012, Shiozaki2014, bernevig2017, Volovik2009, FuAndo2015, Sato2017, senthilmarston1999, Read2000} typically encounter the notorious non-Hermitian (NH) skin effect in open systems interacting with a bath or the environment~\cite{Torres2020, Ghatak2019, Bergholtz2021, Esaki2011, Liang2013, Yao2018Aug, Yao2018Sep, Gong2018, Kawabata2018, Shen2018, Kunst2018, Kawabata2019, Zhou2019, Yokomizo2019, kawabata2019b, Lee2019, Okuma2020, Zhang2020Sep, Borgnia2020, Wu2020, Sun2021, Zhu2021, PanigraphiNH2022, Manna2023}. It corresponds to an accumulation of all the right and left eigenvectors at opposite ends of the system with open boundary conditions. Even though topological invariants can be defined in NH systems their signatures on the the boundary modes get masked by the NH skin effect. Theoretically, this obstacle can be bypassed through the bi-orthogonal bulk-boundary correspondence, devoid of any NH skin effect~\cite{Kunst2018}. Nevertheless, any direct manifestation of the topological invariants on experimentally measurable quantities even in an NH toy topological model remains illusive~\cite{CosmaNHtrans2024}, leaving aside their actual measurements in open quantum systems. In this quest, two-dimensional NH topological models stand as cornerstones, since in the Hermitian or closed systems their topological invariants can be computed and measured in six-terminal Hall-bar setups~\cite{Klitzing1980, Konig2007, Knez2011, ZhongFang2010, QKXu2013, Chang2015, Banerjee2018, Kasahara2018, Srivastav2019, Breton2022}.

\begin{figure}[t!]
\includegraphics[width=1.00\linewidth]{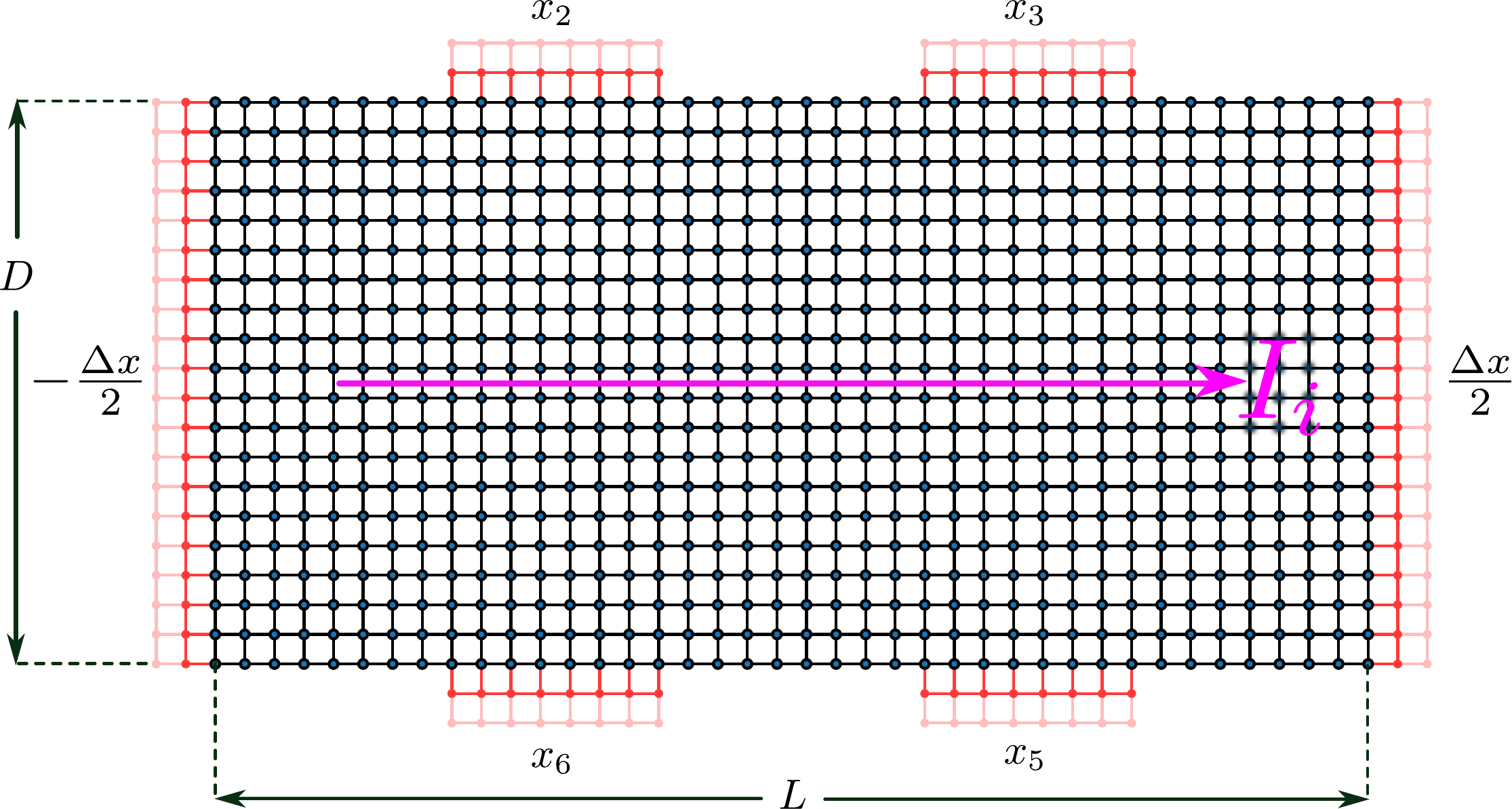}
\caption{Six-terminal setup with a rectangular scattering region (black sites) of length $L$ and width $D$ connected to six semi-infinite leads (red sites), for the calculation of electrical ($\sigma_{xy}$), thermal ($\kappa_{xy}$), and spin ($\sigma^{sp}_{xy}$) Hall, and electrical ($G_{xx}$) and thermal ($G^{th}_{xx}$) longitudinal conductivities. For the computation of $\sigma_{xy}$ and $G_{xx}$, an electrical current ($I_{el}$) flows due to a voltage bias ($\Delta x \equiv \Delta V$) between the horizontal leads, and we extract voltages $x_i \equiv V_i$ on the vertical leads with $i=2,3,5,6$. Due to a temperature gradient ($\Delta x \equiv \Delta T$), a thermal current ($I_{th}$) flows between the horizontal leads and temperature develops at the vertical leads with $x_i \equiv T_i$, when we compute $\kappa_{xy}$ and $G^{th}_{xx}$. While calculating $\sigma^{sp}_{xy}$, a spin current ($I_{sp}$) flows between the horizontal leads, subject to a magnetic field bias with $\Delta x \equiv \Delta H$, yielding magnetization at the vertical leads with $x_i \equiv m_i$. The scattering region is maintained at a fixed voltage $V$ (for $\sigma_{xy}$ and $G_{xx}$) or temperature $T$ (for $\kappa_{xy}$ and $G^{th}_{xx}$) or magnetic field $H$ (for $\sigma^{sp}_{xy}$). For all calculations (Figs.~\ref{fig:NHQAHIQSHI}-\ref{fig:disorder}), we set $L=180$ and $D=90$.        
}~\label{fig:setup}
\end{figure}

Here, we consider skin effect-free NH generalization of square lattice models for all five non-trivial Hermitian Altland-Zirnbauer symmetry classes~\cite{Altland1997} that over an extended NH parameter regime feature real eigenvalue spectrum and non-vanishing topological invariants~\cite{SalibDasRoy2023}. Using the Kwant software package~\cite{Groth2014, Fulga2020, SKDasPRB2023THC, SKDasPRB2024THC}, we  numerically compute their electrical ($\sigma_{xy}$), thermal ($\kappa_{xy}$), and spin ($\sigma^{sp}_{xy}$) Hall, and electrical ($G_{xx}$) and thermal ($G^{th}_{xx}$) longitudinal conductivities in six-terminal setups (Fig.~\ref{fig:setup}), to arrive at the following conclusions.

\begin{figure}[t!]
\includegraphics[width=1.00\linewidth]{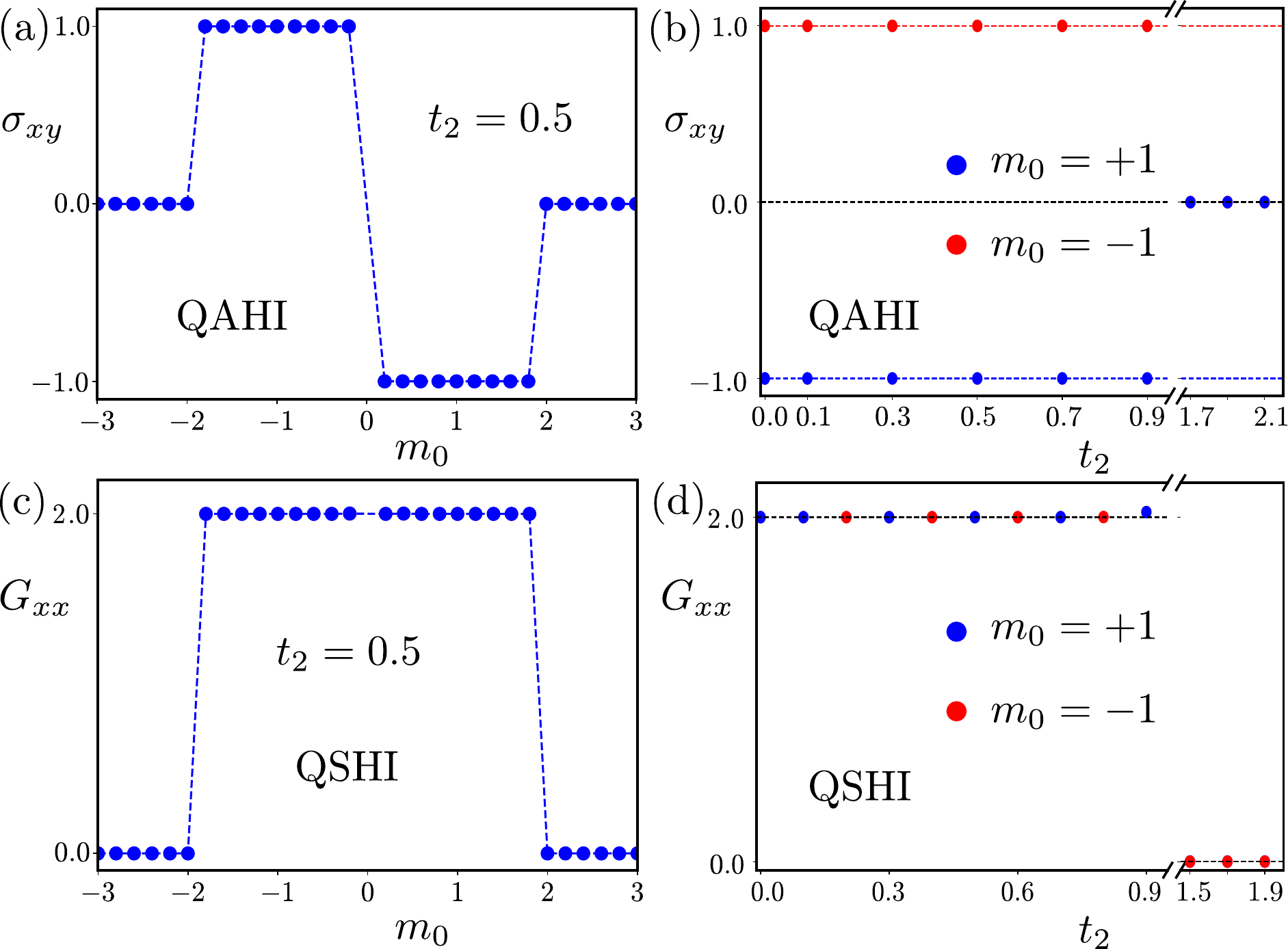}
\caption{Electrical Hall conductivity ($\sigma_{xy}$) of an NH QAHI for (a) a fixed strength of non-Hermiticity ($t_2$) as a function of $m_0$ and (b) two fixed values of $m_0$ with varying $t_2$, showing $\sigma_{xy}= C$ in units of $e^2/h$, where $C$ is the first Chern number. Electrical longitudinal conductivity ($G_{xx}$) of an NH QSHI for (c) a fixed $t_2$ as a function of $m_0$ and (d) two fixed values of $m_0$ with varying $t_2$, showing $G_{xx}= C_{sp}$ (in units of $e^2/h$), where $ C_{sp}$ is the spin Chern number. For $t_2>1$, $\sigma_{xy}=G_{xx}=0$.        
}~\label{fig:NHQAHIQSHI}
\end{figure}

\begin{figure}[t!]
\includegraphics[width=1.00\linewidth]{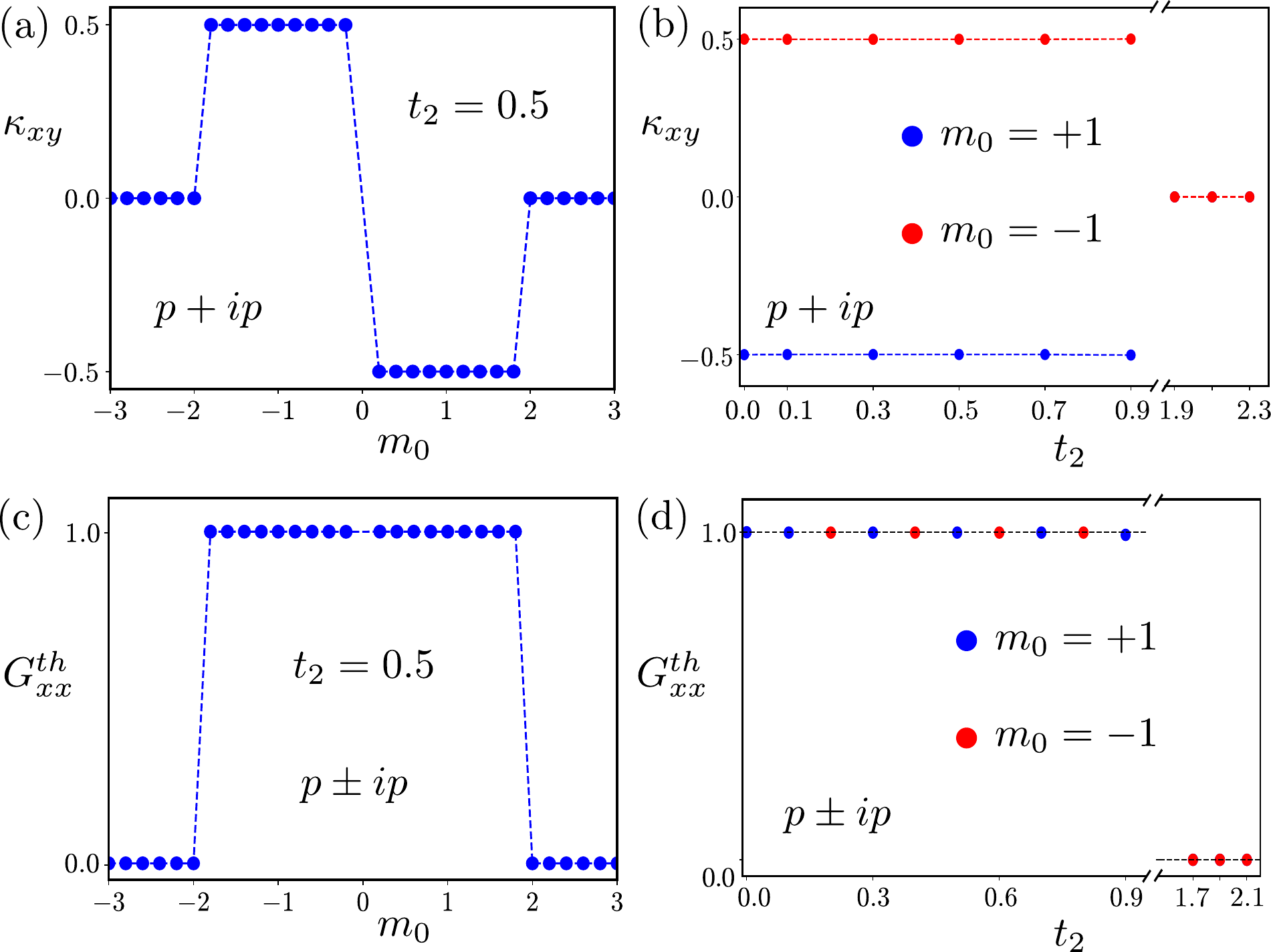}
\caption{Thermal Hall conductivity ($\kappa_{xy}$) of an NH $p+ip$ paired state for (a) a fixed $t_2$ as a function of $m_0$ and (b) two fixed values of $m_0$ with varying $t_2$, showing half-quantization $\kappa_{xy}= C/2$ (in units of $\kappa_0$). Thermal longitudinal conductivity ($G^{th}_{xx}$) of an NH $p \pm ip$ superconductor for (c) a fixed $t_2$ with varying $m_0$ and (d) two fixed values of $m_0$ with varying $t_2$, showing $G^{th}_{xx}= C_{sp}/2$ (in units of $\kappa_0$). For $t_2>1$, $\kappa_{xy}=G^{th}_{xx}=0$.       
}~\label{fig:pwave}
\end{figure}

\begin{figure}[t!]
\includegraphics[width=1.00\linewidth]{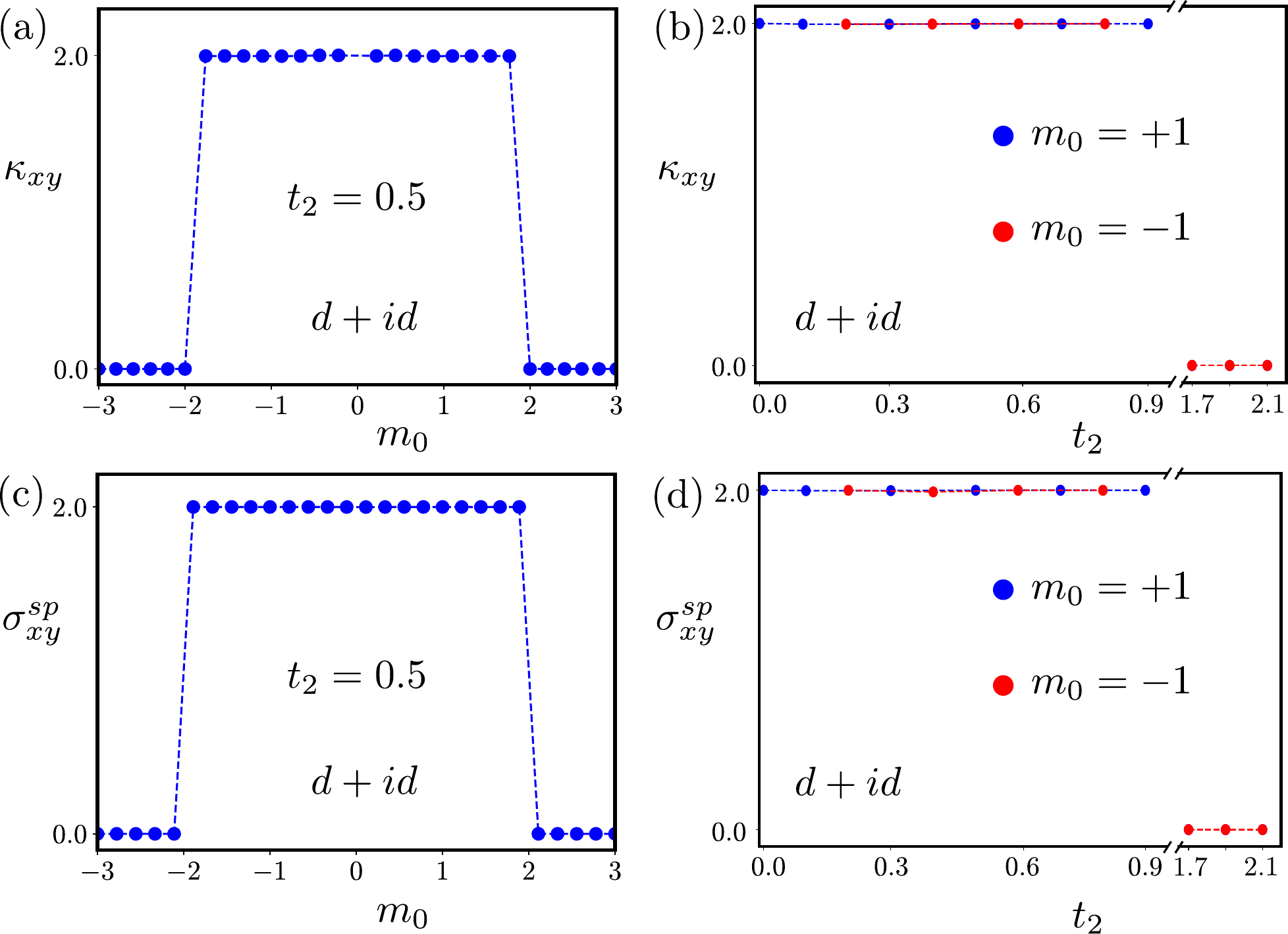}
\caption{Thermal Hall conductivity ($\kappa_{xy}$) of an NH $d+id$ pairing for (a) a fixed $t_2$ with varying $m_0$ and (b) two fixed values of $m_0$ with varying $t_2$, showing $\kappa_{xy}= C$ (in units of $\kappa_0$), where $C=2$. Its spin Hall conductivity ($\sigma^{sp}_{xy}$) for (c) a fixed $t_2$ as a function of $m_0$ and (d) two fixed values of $m_0$ with varying $t_2$ shows $\sigma^{sp}_{xy}= C$ (in units of $\sigma^{sp}_0$). For $t_2>1$, $\kappa_{xy}=\sigma^{sp}_{xy}=0$.       
}~\label{fig:dwave}
\end{figure}

\begin{figure*}[t!]
\includegraphics[width=1.00\linewidth]{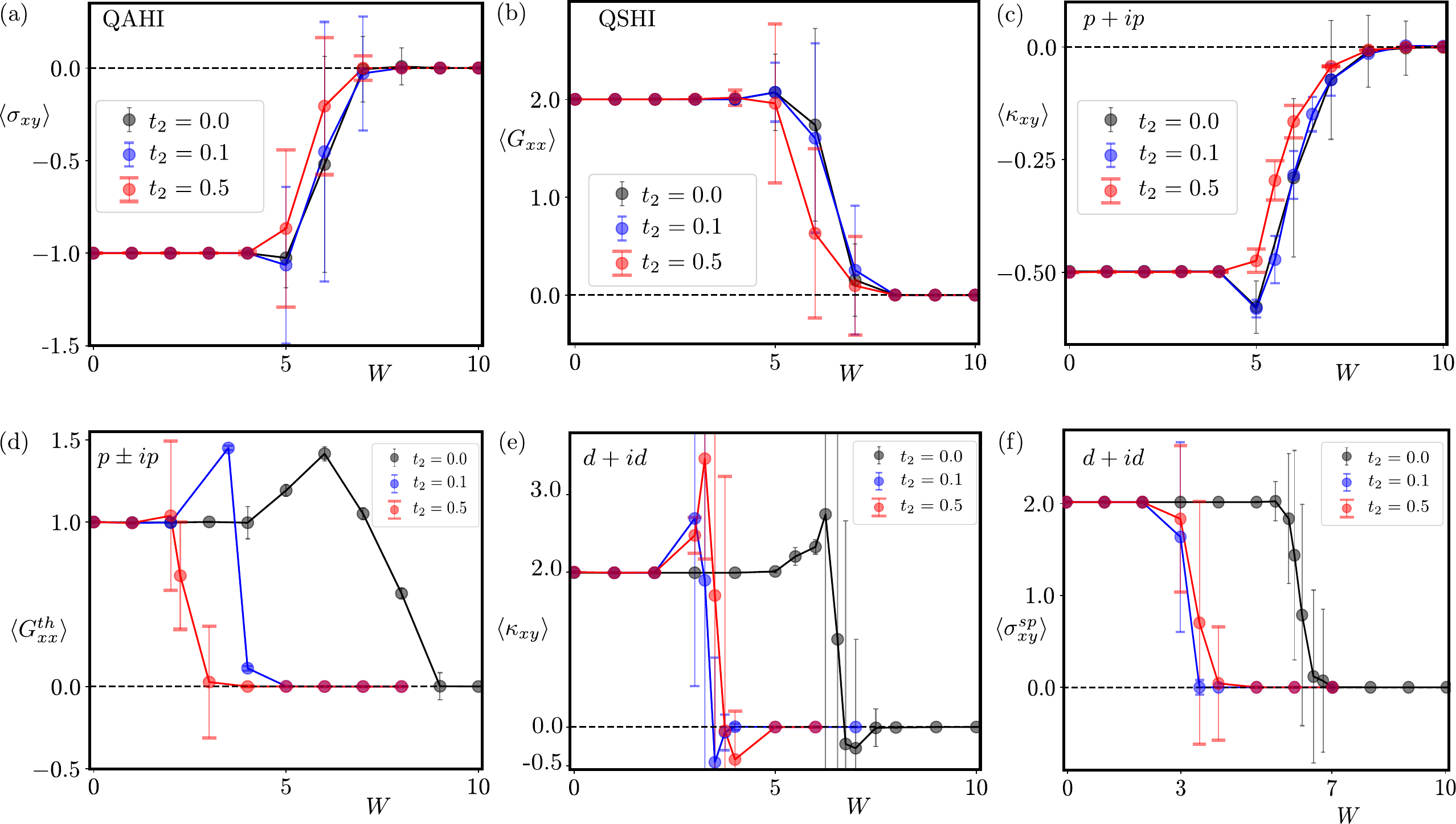}
\caption{Disorder averaged (a) electrical Hall conductivity $\langle \sigma_{xy} \rangle$ in QAHI, (b) electrical longitudinal conductivity $\langle G_{xx} \rangle$ in QSHI, (c) thermal Hall conductivity $\langle \kappa_{xy} \rangle$ in $p+ip$ paired state, (d) thermal longitudinal conductivity $\langle G^{th}_{xx} \rangle$ in $p \pm ip$ superconductor, and (e) $\langle \kappa_{xy} \rangle$ and (f) spin Hall conductivity $\langle \sigma^{sp}_{xy} \rangle$ in $d+id$ paired state in Hermitian ($t_2=0$) and NH ($t_2=0.1,0.5$) systems. The error bars correspond to the saturated (after averaging over a large number of disorder realizations) standard deviations~\cite{SM}.          
}~\label{fig:disorder}
\end{figure*}

The NH quantum anomalous Hall insulator (QAHI) and quantum spin Hall insulator (QSHI) belonging to class A and class AII in the Hermitian limit, respectively, show quantized $\sigma_{xy}$ and $G_{xx}$ in units of $e^2/h$ (Fig.~\ref{fig:NHQAHIQSHI}). Here $e$ ($h$) is the electrical charge (Planck's constant). By contrast, the NH $p+ip$ ($p \pm ip$) paired state, belonging to class D (class DIII) in Hermitian systems, supports half-quantized $\kappa_{xy}$ ($G^{th}_{xx}$) in units of $\kappa_0=\pi^2 k^2_{\rm B} T/(3h)$ as the temperature $T \to 0$, where $k_{\rm B}$ is the Boltzmann constant (Fig.~\ref{fig:pwave}). Finally, the NH $d+id$ paired state, falling in class C in the Hermitian limit, accommodates quantized $\kappa_{xy}$ (in units of $\kappa_0$) and $\sigma^{sp}_{xy}$ (in units of $\sigma^{sp}_0=\hbar/(8\pi)$). See Fig.~\ref{fig:dwave}. These conclusions hold in the entire topological regime in a clean and weakly disordered (due to pointlike random charge impurities) systems. In the strong disorder regime, all these responses vanish. However, with increasing $t_2$ or non-Hermiticity in the system, the onset of vanishing topological responses occur at weaker disorder (Fig.~\ref{fig:disorder}). This phenomenon takes place via quantum phase transitions at finite disorder, but the associated critical disorder strength ($W_c$) can only be pinned from numerical simulations in sufficiently large systems, which is beyond the scope of our numerical resources.

\emph{Model}.~For all the gapped topological phases belonging to any one of the five non-trivial Altland-Zirnbauer symmetry classes in two dimensions, the model Hamiltonian on a square lattice take the following universal form 
\begin{equation}~\label{eq:AZUniversal}
\hspace{-0.275cm}
H= \alpha \sum_{\vec{k}} \Psi^\dagger_{\vec{k}} \bigg( \sum^{3}_{j=1} d_j(\vec{k}) \Gamma_j \bigg) \Psi_{\vec{k}} 
\equiv \alpha \sum_{\vec{k}} \Psi^\dagger_{\vec{k}} {\mathcal H}(\vec{k}) \Psi_{\vec{k}}. 
\end{equation} 
The internal structure of the spinor $\Psi_{\vec{k}}$ with momentum $\vec{k}=(k_x,k_y)$ and representation of the Hermitian $\Gamma$ matrices depend on the symmetry class. They always satisfy the anticommuting Clifford algebra $\{\Gamma_j, \Gamma_k \}=2 \delta_{jk}$. The third component of the $\vec{d}(\vec{k})$-vector takes the form $d_3(\vec{k})= m_0-t_0 \left[ \cos(k_x a) + \cos(k_y a) \right]$, and unless otherwise stated $d_1 (\vec{k}) = t_1 \sin(k_x a)$ and $d_2 (\vec{k}) = t_1 \sin(k_y a)$, where $a$ is the lattice spacing. Within the topological regime ($|m_0/t_0| <2$), for topological insulators (superconductors) $d_3(\vec{k})$ features a band inversion (Fermi surface) near the $\Gamma=(0,0)$ and ${\rm M}=(1,1)\pi/a$ points of the Brillouin zone (BZ) when $0<m_0/t_0 <2$ and $-2<m_0/t_0<0$, respectively, with $\alpha=1$ ($\alpha=1/2$ due to the Nambu doubling). Throughout, we set $t_1=t_0=1$.

Note that $d_3(\vec{k})$ and $\Gamma_3$ do not break any crystal symmetry, and ${\mathcal H}_\perp(\vec{k})=d_1(\vec{k}) \Gamma_1 +  d_2(\vec{k}) \Gamma_2$ is also invariant under all the crystal symmetries. So, ${\mathcal H}(\vec{k})$ transforms under the trivial $A_{1g}$ representation. Then the anti-Hermitian operator ${\mathcal H}_{\rm AH}(\vec{k})=\Gamma_3 {\mathcal H}_\perp(\vec{k})/t_1$ does not break any crystallographic symmetry either. Finally, we introduce the desired NH topological operator 
\begin{equation}~\label{eq:NHGeneral}
{\mathcal H}_{\rm NH}(\vec{k})= {\mathcal H}(\vec{k}) + t_2 {\mathcal H}_{\rm AH}(\vec{k})
\end{equation}   
that is symmetric under all crystal symmetries, such as reflections, four-fold rotations, and inversion. Hence, it is devoid of any NH skin effect for all parameter values~\cite{SalibDasRoy2023}. Here, $t_2$ determines the strength of non-Hermiticity in the system. Eigenvalues of ${\mathcal H}_{\rm NH}(\vec{k})$ are $\pm E_{\rm NH}(\vec{k})$, where
\begin{equation}~\label{eq:NHEigen}
E_{\rm NH}(\vec{k}) =\left[ (t^2_1-t^2_2) \{ d^2_1(\vec{k}) + d^2_2(\vec{k}) \} + d^2_3(\vec{k}) \right]^{1/2}.
\end{equation}
They are purely real when $t_2<t_1$. Then ${\mathcal H}_{\rm NH}(\vec{k})$ fosters non-trivial topological invariants, when $|m_0/t_0|<2$. Our construction from Eq.~\eqref{eq:NHGeneral} is applicable to any lattice systems, such as the triangular one, where the parent Hamiltonian also assumes the universal form of Eq.~\eqref{eq:AZUniversal}~\cite{Slager2012}. The resulting NH operator respects the corresponding crystal symmetries, such as three-fold rotations.

\emph{NH QAHI}.~For $\Gamma_j =\tau_j$ and $\Psi^\top_{\vec{k}}=(c_{+}, c_{-})(\vec{k})$, where $c_{\tau}(\vec{k})$ is the fermionic annihilation operator on orbital with parity eigenvalue $\tau=\pm$ and momentum $\vec{k}$, we find the NH incarnation of the Qi-Wu-Zhang model for QAHI~\cite{Qi2006}. The Pauli matrices $\{ \tau_j \}$ act on the orbital index. Then the NH operator takes the form ${\mathcal H}_{\rm NH}(\vec{k}) = \sum^3_{j=1}d^{\rm NH}_j(\vec{k}) \tau_j$, with $d^{\rm NH}_1(\vec{k})=t_1 \sin(k_x a)-i t_2 \sin(k_y a)$, $d^{\rm NH}_2(\vec{k})=t_1 \sin(k_y a)+i t_2 \sin(k_x a)$, and $d^{\rm NH}_3(\vec{k})=d_3(\vec{k})$. The first Chern number of ${\mathcal H}_{\rm NH}(\vec{k})$ reads~\cite{Thouless1982, SalibDasRoy2023}  
\begin{equation}~\label{eq:chernnumber}
C= \int_{\rm BZ} \frac{d^2 \vec{k}}{4\pi} \left[ \partial_{k_x} \hat{\vec{d}}^{\rm NH}(\vec{k}) \times \partial_{k_y} \hat{\vec{d}}^{\rm NH}(\vec{k}) \right] \cdot \hat{\vec{d}}^{\rm NH}(\vec{k}),
\end{equation}
where $\hat{\vec{d}}^{\rm NH}(\vec{k})=\vec{d}^{\rm NH}(\vec{k})/\sqrt{[\vec{d}^{\rm NH}(\vec{k})]^2}$. The momentum integral is performed over the first BZ. For any $|t_2|<1$, we find $C=-1$ ($+1$) for $0<m_0/t_0<2$ ($-2<m_0/t_0<0$). Next, we discuss the ramification of non-trivial $C$ on the quantized $\sigma_{xy}$, computed in a six-terminal setup (Fig.~\ref{fig:setup}), detailed in the Supplemental Material (SM)~\cite{SM}, with all the numerical codes available on Zenodo~\cite{zenodoTSCSanjib}.

A voltage gradient is applied between two horizontal leads, yielding a longitudinal electrical current ($I_{el}$) between them. The transverse leads, serving as the voltage probes, carry no electrical current. The current-voltage relation is then given by ${\bf I}_{el}= {\bf G}_{el} {\bf V}$, with ${\bf V}^\top=(-\Delta V/2,V_2,V_3,\Delta V/2,V_5,V_6)$ and ${\bf I}^\top_{el}=(I_{el},0,0,-I_{el},0,0)$. Upon computing the conductance matrix ${\bf G}_{el}$ using Kwant~\cite{Groth2014}, containing only the transmission blocks of the scattering matrix, we extract different voltages from the current-voltage relation. Subsequently, we compute the transverse electrical resistance $R^{el}_{xy}=(V_2+V_3-V_5-V_6)/(2 I_{el})$, and find $\sigma_{xy}=1/R^{el}_{xy}=C$ (in units of $e^2/h$). See Figs.~\ref{fig:NHQAHIQSHI}(a) and~\ref{fig:NHQAHIQSHI}(b), and SM~\cite{SM}.

\emph{NH QSHI}.~An NH generalization of the Bernevig-Hughes-Zhang model for QSHI is realized with ${\boldsymbol \Gamma}=(\Gamma_{01}, \Gamma_{32}, \Gamma_{03})$, where $\Gamma_{\mu \nu}=\sigma_\mu \tau_\nu$ and the Pauli matrices $\{ \sigma_\mu \}$ operate on the spin index~\cite{BHZ2006}. The four-component spinor reads as $\Psi^\top_{\vec{k}}=(c_+^{\uparrow}, c_-^{\uparrow}, c_+^{\downarrow}, c_-^{\downarrow})(\vec{k})$, where $c_\tau^{\sigma}(\vec{k})$ is the fermionic annihilation operator with parity $\tau=\pm$, spin projection $\sigma=\uparrow, \downarrow$ in the $z$ direction, and momentum $\vec{k}$. Following the same steps, highlighted for the NH QAHI model, we compute the Chern number for individual spin components, given by $C_\uparrow$ and $C_\downarrow$. Within the topological regime, we find that the total Chern number of the model is $C_{total}=C_\uparrow + C_\downarrow =0$. Nevertheless, we can define an invariant, the spin Chern number $C_{sp}=|C_\uparrow - C_\downarrow|$ which is non-trivial in the entire topological regime ($|m_0/t_0|<2$ and $|t_2|<1$), given by $C_{sp}=2$.    
    
The non-trivial $C_{sp}$ leaves its footprint on the quantized $G_{xx}$ for which we employ the same six-terminal arrangement, previously discussed for the NH QAHI. In this case, $\sigma_{xy}=0$ as $C_{total}=0$ therein. But, from the same current-voltage relationship, we now compute the longitudinal electrical resistance $R_{xx}=(V_3-V_2)/I_{el}=(V_5-V_6)/I_{el}$, in turn yielding $G_{xx} = R^{-1}_{xx} = C_{sp}$ (in units of $e^2/h$), as shown in Figs.~\ref{fig:NHQAHIQSHI}(c) and~\ref{fig:NHQAHIQSHI}(d), due to the associated counter-propagating helical edge modes.

\emph{NH $p+ip$ pairing}.~For an NH $p + ip$ paired state among spinless or spin-polarized fermions $\Psi^\top_{\vec{k}}=(c_{\vec{k}}, c^\dagger_{-\vec{k}})$, where $c_{\vec{k}}$ ($c^\dagger_{\vec{k}}$) is the fermionic annihilation (creation) operator with momentum $\vec{k}$. The Pauli matrices {\color{blue}$\{ \tau_j \}$} operate on the Nambu or particle-hole index~\cite{Read2000}. The effective single-particle NH Bogoliubov de-Gennes operator ${\mathcal H}_{\rm NH}(\vec{k})$ has the topological invariant $C$ [Eq.~\eqref{eq:chernnumber}]. It manifests via a half-quantized $\kappa_{xy}$. Now, a thermal current ($I_{th}$) flows between two horizontal leads, held at fixed but different temperatures (Fig.~\ref{fig:setup}). Four vertical or transverse leads serve as the temperature probe, carry no thermal current. The thermal current-temperature relation is a matrix equation ${\bf I}_{th}= {\bf A}_{th} {\bf T}$, where ${\bf I}^\top_{th}=(I_{th},0,0,-I_{th},0,0)$ and ${\bf T}^\top=(-\Delta T/2,T_2,T_3,\Delta T/2,T_5,T_6)$. The elements of ${\bf A}_{th}$ are 
\begin{equation}~\label{eq:currtemp}
A_{th,ij} =  \int_0^\infty \frac{E^2}{T} \left( -\frac{\partial f(E, T)}{\partial E} \right)\left[\delta_{ij} \mu_j-  {\rm Tr}({\bf t}_{ij}^\dagger {\bf t}_{ij}) \right] dE,
\end{equation}
where $f(E,T)=1/(1+\exp{[E/(k_{\rm B}T)]})$ is the Fermi-Dirac distribution function, $\mu_j$ denotes the number of propagating modes in the $j$th lead,  ${\bf t}_{ij}$ is the transmission part of the scattering matrix between the leads $i$ and $j$, and the trace (Tr) is taken over the conducting channels. From ${\bf A}_{th}$, we extract temperatures at various leads. The transverse thermal resistance is then given by $R^{th}_{xy}=(T_2+T_3-T_5-T_6)/(2I_{th})$~\cite{Fulga2020, SKDasPRB2023THC, SKDasPRB2024THC, grapheneTHC2011}, from which we compute $\kappa_{xy}= 1/R^{th}_{xy}$ at $T=0.01$, showing $k_{xy}=\alpha C$ in units of $\kappa_0$ (Figs.~\ref{fig:pwave}(a) and~\ref{fig:pwave}(b), and SM~\cite{SM}), where $\alpha=1/2$ accounts for the Nambu doubling. Thus, model NH $p+ip$ paired state features a half-quantized $\kappa_{xy}$ in the entire topological regime.

\emph{NH $p \pm ip$ pairing}.~This paired state occurs among spin-1/2 fermions, with $p + ip$ and $p-ip$ pairing symmetries for opposite spin projections. The associated four-component spinor is $\Psi_{\vec{k}}=(\psi_{\vec{k}}, \sigma_2 \psi^\star_{-\vec{k}})$ with $\psi_{\vec{k}}=(c_\uparrow, c_\downarrow)(\vec{k})$, where $c_\sigma(\vec{k})$ is the fermionic annihilation operator with spin projection $\sigma=\uparrow, \downarrow$ and momentum $\vec{k}$. The involved $\Gamma$ matrices are $\Gamma_1=\sigma_0 \tau_1$, $\Gamma_2=\sigma_3 \tau_2$, and $\Gamma_3=\sigma_0 \tau_3$. Similar to the situation in NH QSHI, $C_{total}=0$ but $C_{sp}=2$ for the NH $p \pm ip$ paired. We numerically confirm that now $\kappa_{xy}=0$. Nevertheless, from the longitudinal thermal resistance $R^{th}_{xx}=(T_3 -T_2)/I_{th}=(T_6 -T_5)/I_{th}$, we find that $G^{th}_{xx}=(R^{th}_{xx})^{-1}=\alpha C_{sp} \kappa_0$ with $\alpha=1/2$, as shown in Figs.~\ref{fig:pwave}(c) and~\ref{fig:pwave}(d).

\emph{NH $d+id$ pairing}.~In the same Nambu-doubled spinor basis, a spin-singlet NH $d+id$ paired state is realized with $d_1(\vec{k})=t_1 [\cos(k_x a)-\cos(k_y a)]$, $d_2(\vec{k})=t_1 \sin(k_x a) \sin(k_y a)$, and $\Gamma_j=\sigma_0 \tau_j$~\cite{Read2000, Laughlin1998}. Within the topological regime ($|m_0/t_0|<2$ and $|t_2|<1$), the Chern number of the corresponding NH operator for each spin projection is $C=2$, and thus its total Chern number is $2C=4$. This NH paired state accommodates $\kappa_{xy}=C$ (in units of $\kappa_0$), as shown in Fig.~\ref{fig:dwave}(a) and~\ref{fig:dwave}(b). The spin-charge separation allows this NH paired state to harbor quantized spin Hall conductivity ($\sigma^{sp}_{xy}$), for which a spin current ($I_{sp}$) passes through the scattering region due to a difference in the magnetic field bias between the two horizontal leads. Then magnetization ($m_j$) develops in the vertical leads, which we compute from the spin current-magnetization matrix relation ${\bf I}_{sp}={\bf G}_{sp} {\bf M}$, where ${\bf I}^\top_{sp}=(I_{sp},0,0,-I_{sp},0,0)$ and ${\bf M}^\top=(-\Delta H/2, m_2, m_3, \Delta H/2, m_5, m_6)$, and ${\bf G}_{sp}$ is the spin conductance matrix, extracted using Kwant~\cite{SKDasPRB2024THC}. From its solutions, we compute the spin Hall resistance $R^{sp}_{xy}=(m_2+m_3-m_5-m_6)/(2 I_{sp})$, yielding $\sigma^{sp}_{xy}=\left( R^{sp}_{xy} \right)^{-1}=C$ (in units of $\sigma^{sp}_0$), see Figs.~\ref{fig:dwave}(c) and~\ref{fig:dwave}(d).

\emph{Disorder}.~Finally, we unfold the effects of disorder on all the topological responses, discussed so far in clean NH systems. We consider only pointlike charge impurities, the dominant source of elastic scattering in any real material. In the above NH systems, we introduce the terms $V(\vec{r}) \tau_0$, $V(\vec{r}) \sigma_0 \tau_0$, $V(\vec{r}) \tau_3$, $V(\vec{r}) \sigma_0 \tau_3$, and $V(\vec{r}) \sigma_0 \tau_3$, respectively, where $V(\vec{r})$ is uniformly and randomly distributed within the range $[-W/2, W/2]$ at each site of the scattering region and $W$ denotes the strength of disorder. The results are shown in Fig.~\ref{fig:disorder}. All the topological transport quantities (disorder averaged) retain robust (half-)quantized values in the weak disorder regime, while they all vanish in the strong disorder regime. Furthermore, with the introduction of non-Hermiticity in the system as $t_1 \to \sqrt{t^2_1-t^2_2}$ [Eq.~\eqref{eq:NHEigen}], fermions become weakly dispersive and the latter event occurs at weaker disorder, which can be seen by comparing the results for $t_2=0$ and $0.5$. However, even $L=2D=180$ systems (Fig.~\ref{fig:setup}) turn out to be insufficient to compare the results for different finite $t_2$ values. Numerical simulations in larger systems become too time expensive for us to pursue at this stage. Nevertheless, most of the results from Fig.~\ref{fig:disorder} strongly suggest that with increasing $t_2$, the disappearance of the topological responses occurs at weaker disorder. It can be seen by comparing the results for $t_2=0.1$ and $0.5$. Then the transport quantities deviate from their (half-)quantized values at weaker disorder for $t_2=0.5$ than for $t_2=0.1$, except for the $d+id$ paired state, for which the curves are very close to each other. The prominent finite size effects in the $d+id$ paired state due to longer range hopping encoded in $d_2(\vec{k})$ is confirmed by numerically computing disorder averaged $\sigma^{sp}_{xy}$ in a smaller $L=2D=120$ system~\cite{SM}.

\emph{Discussions}.~Here we identify a family of NH operators, devoid of NH skin effects, for planar topological insulators and superconductors that manifests the topological invariant through (half-)quantized electrical, thermal, and spin transport quantities, closely mimicking the ones previously found in Hermitian systems~\cite{SKDasPRB2023THC, SKDasPRB2024THC}. For any $|t_2|>1$, when the bulk topological invariants ($C$ and $C_{sp}$) vanish, there is no quantization of any transport quantity. In the SM, we analytically show that the topological bound state exists only when $t_2<1$~\cite{SM}. However, in order to observe vanishing transport responses we need to set $t_2$ to be slightly bigger than unity to bypass finite size effects [Figs.~\ref{fig:NHQAHIQSHI}-\ref{fig:dwave}]. In model NH operators, displaying the NH skin effect, besides the topological edge modes all the left and right eigenvectors reside near the opposite edges of the scattering region by definition, ruining any (half-)quantized transport responses therein. Even though the requisite six-terminal Hall bar arrangement is well-developed by now, controlled synthesis of NH quantum crystals remains far from reality. Nevertheless, optical lattices of neutral atoms constitute a promising testbed for our theoretical predictions, where a plethora of topological band engineering protocols has been proposed~\cite{Discussion1} and realized~\cite{Discussion2}. Simplicity of our construction, in which the NH operators result from nearest-neighbor hopping modulations, makes them achievable on such highly tunable platforms that also harbor superfluids (charge-neutral superconductors). In such systems, the Hall conductivity can be obtained from the ``heating effect"~\cite{Discussion3}, for example, which can also be employed to measure quantized longitudinal transports.

\emph{Acknowledgments.}~S.K.D.\ was supported by the Startup Grant of B.R.\ from Lehigh University. B.R.\ was supported by NSF CAREER Grant No.\ DMR-2238679. Portions of this research were conducted on Lehigh University's Research Computing infrastructure partially supported by NSF Award No.~2019035. We thank Vladimir Juri\v{c}i\'c for critical reading of the manuscript.

\bibliography{Ref_NHTransport} 

\begin{thebibliography}{71}%
\makeatletter
\providecommand \@ifxundefined [1]{%
 \@ifx{#1\undefined}
}%
\providecommand \@ifnum [1]{%
 \ifnum #1\expandafter \@firstoftwo
 \else \expandafter \@secondoftwo
 \fi
}%
\providecommand \@ifx [1]{%
 \ifx #1\expandafter \@firstoftwo
 \else \expandafter \@secondoftwo
 \fi
}%
\providecommand \natexlab [1]{#1}%
\providecommand \enquote  [1]{``#1''}%
\providecommand \bibnamefont  [1]{#1}%
\providecommand \bibfnamefont [1]{#1}%
\providecommand \citenamefont [1]{#1}%
\providecommand \href@noop [0]{\@secondoftwo}%
\providecommand \href [0]{\begingroup \@sanitize@url \@href}%
\providecommand \@href[1]{\@@startlink{#1}\@@href}%
\providecommand \@@href[1]{\endgroup#1\@@endlink}%
\providecommand \@sanitize@url [0]{\catcode `\\12\catcode `\$12\catcode
  `\&12\catcode `\#12\catcode `\^12\catcode `\_12\catcode `\%12\relax}%
\providecommand \@@startlink[1]{}%
\providecommand \@@endlink[0]{}%
\providecommand \url  [0]{\begingroup\@sanitize@url \@url }%
\providecommand \@url [1]{\endgroup\@href {#1}{\urlprefix }}%
\providecommand \urlprefix  [0]{URL }%
\providecommand \Eprint [0]{\href }%
\providecommand \doibase [0]{https://doi.org/}%
\providecommand \selectlanguage [0]{\@gobble}%
\providecommand \bibinfo  [0]{\@secondoftwo}%
\providecommand \bibfield  [0]{\@secondoftwo}%
\providecommand \translation [1]{[#1]}%
\providecommand \BibitemOpen [0]{}%
\providecommand \bibitemStop [0]{}%
\providecommand \bibitemNoStop [0]{.\EOS\space}%
\providecommand \EOS [0]{\spacefactor3000\relax}%
\providecommand \BibitemShut  [1]{\csname bibitem#1\endcsname}%
\let\auto@bib@innerbib\@empty
\bibitem [{\citenamefont {Hasan}\ and\ \citenamefont {Kane}(2010)}]{Hasan2010}%
  \BibitemOpen
  \bibfield  {author} {\bibinfo {author} {\bibfnamefont {M.~Z.}\ \bibnamefont
  {Hasan}}\ and\ \bibinfo {author} {\bibfnamefont {C.~L.}\ \bibnamefont
  {Kane}},\ }\bibfield  {title} {\bibinfo {title} {{Colloquium: Topological
  insulators}},\ }\href {https://doi.org/10.1103/RevModPhys.82.3045} {\bibfield
   {journal} {\bibinfo  {journal} {Rev. Mod. Phys.}\ }\textbf {\bibinfo
  {volume} {82}},\ \bibinfo {pages} {3045} (\bibinfo {year}
  {2010})}\BibitemShut {NoStop}%
\bibitem [{\citenamefont {Qi}\ and\ \citenamefont {Zhang}(2011)}]{Qi2011}%
  \BibitemOpen
  \bibfield  {author} {\bibinfo {author} {\bibfnamefont {X.-L.}\ \bibnamefont
  {Qi}}\ and\ \bibinfo {author} {\bibfnamefont {S.-C.}\ \bibnamefont {Zhang}},\
  }\bibfield  {title} {\bibinfo {title} {{Topological insulators and
  superconductors}},\ }\href {https://doi.org/10.1103/RevModPhys.83.1057}
  {\bibfield  {journal} {\bibinfo  {journal} {Rev. Mod. Phys.}\ }\textbf
  {\bibinfo {volume} {83}},\ \bibinfo {pages} {1057} (\bibinfo {year}
  {2011})}\BibitemShut {NoStop}%
\bibitem [{\citenamefont {Chiu}\ \emph {et~al.}(2016)\citenamefont {Chiu},
  \citenamefont {Teo}, \citenamefont {Schnyder},\ and\ \citenamefont
  {Ryu}}]{Chiu2016}%
  \BibitemOpen
  \bibfield  {author} {\bibinfo {author} {\bibfnamefont {C.-K.}\ \bibnamefont
  {Chiu}}, \bibinfo {author} {\bibfnamefont {J.~C.~Y.}\ \bibnamefont {Teo}},
  \bibinfo {author} {\bibfnamefont {A.~P.}\ \bibnamefont {Schnyder}},\ and\
  \bibinfo {author} {\bibfnamefont {S.}~\bibnamefont {Ryu}},\ }\bibfield
  {title} {\bibinfo {title} {{Classification of topological quantum matter with
  symmetries}},\ }\href {https://doi.org/10.1103/RevModPhys.88.035005}
  {\bibfield  {journal} {\bibinfo  {journal} {Rev. Mod. Phys.}\ }\textbf
  {\bibinfo {volume} {88}},\ \bibinfo {pages} {035005} (\bibinfo {year}
  {2016})}\BibitemShut {NoStop}%
\bibitem [{\citenamefont {Kane}\ and\ \citenamefont
  {Mele}(2005)}]{kanemele2006}%
  \BibitemOpen
  \bibfield  {author} {\bibinfo {author} {\bibfnamefont {C.~L.}\ \bibnamefont
  {Kane}}\ and\ \bibinfo {author} {\bibfnamefont {E.~J.}\ \bibnamefont
  {Mele}},\ }\bibfield  {title} {\bibinfo {title} {{${Z}_{2}$ Topological Order
  and the Quantum Spin Hall Effect}},\ }\href
  {https://doi.org/10.1103/PhysRevLett.95.146802} {\bibfield  {journal}
  {\bibinfo  {journal} {Phys. Rev. Lett.}\ }\textbf {\bibinfo {volume} {95}},\
  \bibinfo {pages} {146802} (\bibinfo {year} {2005})}\BibitemShut {NoStop}%
\bibitem [{\citenamefont {{Bernevig}}\ \emph {et~al.}(2006)\citenamefont
  {{Bernevig}}, \citenamefont {{Hughes}},\ and\ \citenamefont
  {{Zhang}}}]{BHZ2006}%
  \BibitemOpen
  \bibfield  {author} {\bibinfo {author} {\bibfnamefont {B.~A.}\ \bibnamefont
  {{Bernevig}}}, \bibinfo {author} {\bibfnamefont {T.~L.}\ \bibnamefont
  {{Hughes}}},\ and\ \bibinfo {author} {\bibfnamefont {S.-C.}\ \bibnamefont
  {{Zhang}}},\ }\bibfield  {title} {\bibinfo {title} {{Quantum Spin Hall Effect
  and Topological Phase Transition in HgTe Quantum Wells}},\ }\href
  {https://doi.org/10.1126/science.1133734} {\bibfield  {journal} {\bibinfo
  {journal} {Science}\ }\textbf {\bibinfo {volume} {314}},\ \bibinfo {pages}
  {1757} (\bibinfo {year} {2006})}\BibitemShut {NoStop}%
\bibitem [{\citenamefont {Fu}\ \emph {et~al.}(2007)\citenamefont {Fu},
  \citenamefont {Kane},\ and\ \citenamefont {Mele}}]{FuKaneMele2007}%
  \BibitemOpen
  \bibfield  {author} {\bibinfo {author} {\bibfnamefont {L.}~\bibnamefont
  {Fu}}, \bibinfo {author} {\bibfnamefont {C.~L.}\ \bibnamefont {Kane}},\ and\
  \bibinfo {author} {\bibfnamefont {E.~J.}\ \bibnamefont {Mele}},\ }\bibfield
  {title} {\bibinfo {title} {{Topological Insulators in Three Dimensions}},\
  }\href {https://doi.org/10.1103/PhysRevLett.98.106803} {\bibfield  {journal}
  {\bibinfo  {journal} {Phys. Rev. Lett.}\ }\textbf {\bibinfo {volume} {98}},\
  \bibinfo {pages} {106803} (\bibinfo {year} {2007})}\BibitemShut {NoStop}%
\bibitem [{\citenamefont {Fu}\ and\ \citenamefont {Kane}(2007)}]{Fukane2007}%
  \BibitemOpen
  \bibfield  {author} {\bibinfo {author} {\bibfnamefont {L.}~\bibnamefont
  {Fu}}\ and\ \bibinfo {author} {\bibfnamefont {C.~L.}\ \bibnamefont {Kane}},\
  }\bibfield  {title} {\bibinfo {title} {{Topological insulators with inversion
  symmetry}},\ }\href {https://doi.org/10.1103/PhysRevB.76.045302} {\bibfield
  {journal} {\bibinfo  {journal} {Phys. Rev. B}\ }\textbf {\bibinfo {volume}
  {76}},\ \bibinfo {pages} {045302} (\bibinfo {year} {2007})}\BibitemShut
  {NoStop}%
\bibitem [{\citenamefont {Moore}\ and\ \citenamefont
  {Balents}(2007)}]{moorebalents2007}%
  \BibitemOpen
  \bibfield  {author} {\bibinfo {author} {\bibfnamefont {J.~E.}\ \bibnamefont
  {Moore}}\ and\ \bibinfo {author} {\bibfnamefont {L.}~\bibnamefont
  {Balents}},\ }\bibfield  {title} {\bibinfo {title} {{Topological invariants
  of time-reversal-invariant band structures}},\ }\href
  {https://doi.org/10.1103/PhysRevB.75.121306} {\bibfield  {journal} {\bibinfo
  {journal} {Phys. Rev. B}\ }\textbf {\bibinfo {volume} {75}},\ \bibinfo
  {pages} {121306} (\bibinfo {year} {2007})}\BibitemShut {NoStop}%
\bibitem [{\citenamefont {Roy}(2009)}]{rahulroy2009}%
  \BibitemOpen
  \bibfield  {author} {\bibinfo {author} {\bibfnamefont {R.}~\bibnamefont
  {Roy}},\ }\bibfield  {title} {\bibinfo {title} {{Topological phases and the
  quantum spin Hall effect in three dimensions}},\ }\href
  {https://doi.org/10.1103/PhysRevB.79.195322} {\bibfield  {journal} {\bibinfo
  {journal} {Phys. Rev. B}\ }\textbf {\bibinfo {volume} {79}},\ \bibinfo
  {pages} {195322} (\bibinfo {year} {2009})}\BibitemShut {NoStop}%
\bibitem [{\citenamefont {Ryu}\ \emph {et~al.}(2010)\citenamefont {Ryu},
  \citenamefont {Schnyder}, \citenamefont {Furusaki},\ and\ \citenamefont
  {Ludwig}}]{Ryu2010}%
  \BibitemOpen
  \bibfield  {author} {\bibinfo {author} {\bibfnamefont {S.}~\bibnamefont
  {Ryu}}, \bibinfo {author} {\bibfnamefont {A.~P.}\ \bibnamefont {Schnyder}},
  \bibinfo {author} {\bibfnamefont {A.}~\bibnamefont {Furusaki}},\ and\
  \bibinfo {author} {\bibfnamefont {A.~W.~W.}\ \bibnamefont {Ludwig}},\
  }\bibfield  {title} {\bibinfo {title} {{Topological insulators and
  superconductors: tenfold way and dimensional hierarchy}},\ }\href
  {https://doi.org/10.1088/1367-2630/12/6/065010} {\bibfield  {journal}
  {\bibinfo  {journal} {New J. Phys.}\ }\textbf {\bibinfo {volume} {12}},\
  \bibinfo {pages} {065010} (\bibinfo {year} {2010})}\BibitemShut {NoStop}%
\bibitem [{\citenamefont {Kitaev}(2009)}]{Kitaev2009}%
  \BibitemOpen
  \bibfield  {author} {\bibinfo {author} {\bibfnamefont {A.}~\bibnamefont
  {Kitaev}},\ }\bibfield  {title} {\bibinfo {title} {{Periodic table for
  topological insulators and superconductors}},\ }\href
  {https://doi.org/10.1063/1.3149495} {\bibfield  {journal} {\bibinfo
  {journal} {AIP Conf. Proc.}\ }\textbf {\bibinfo {volume} {1134}},\ \bibinfo
  {pages} {22} (\bibinfo {year} {2009})}\BibitemShut {NoStop}%
\bibitem [{\citenamefont {Schnyder}\ \emph {et~al.}(2008)\citenamefont
  {Schnyder}, \citenamefont {Ryu}, \citenamefont {Furusaki},\ and\
  \citenamefont {Ludwig}}]{Schnyder2008}%
  \BibitemOpen
  \bibfield  {author} {\bibinfo {author} {\bibfnamefont {A.~P.}\ \bibnamefont
  {Schnyder}}, \bibinfo {author} {\bibfnamefont {S.}~\bibnamefont {Ryu}},
  \bibinfo {author} {\bibfnamefont {A.}~\bibnamefont {Furusaki}},\ and\
  \bibinfo {author} {\bibfnamefont {A.~W.~W.}\ \bibnamefont {Ludwig}},\
  }\bibfield  {title} {\bibinfo {title} {{Classification of topological
  insulators and superconductors in three spatial dimensions}},\ }\href
  {https://doi.org/10.1103/PhysRevB.78.195125} {\bibfield  {journal} {\bibinfo
  {journal} {Phys. Rev. B}\ }\textbf {\bibinfo {volume} {78}},\ \bibinfo
  {pages} {195125} (\bibinfo {year} {2008})}\BibitemShut {NoStop}%
\bibitem [{\citenamefont {Fu}(2011)}]{Fu2011}%
  \BibitemOpen
  \bibfield  {author} {\bibinfo {author} {\bibfnamefont {L.}~\bibnamefont
  {Fu}},\ }\bibfield  {title} {\bibinfo {title} {{Topological Crystalline
  Insulators}},\ }\href {https://doi.org/10.1103/PhysRevLett.106.106802}
  {\bibfield  {journal} {\bibinfo  {journal} {Phys. Rev. Lett.}\ }\textbf
  {\bibinfo {volume} {106}},\ \bibinfo {pages} {106802} (\bibinfo {year}
  {2011})}\BibitemShut {NoStop}%
\bibitem [{\citenamefont {Slager}\ \emph {et~al.}(2012)\citenamefont {Slager},
  \citenamefont {Mesaros}, \citenamefont {Juri{\v{c}}i{\'{c}}},\ and\
  \citenamefont {Zaanen}}]{Slager2012}%
  \BibitemOpen
  \bibfield  {author} {\bibinfo {author} {\bibfnamefont {R.-J.}\ \bibnamefont
  {Slager}}, \bibinfo {author} {\bibfnamefont {A.}~\bibnamefont {Mesaros}},
  \bibinfo {author} {\bibfnamefont {V.}~\bibnamefont {Juri{\v{c}}i{\'{c}}}},\
  and\ \bibinfo {author} {\bibfnamefont {J.}~\bibnamefont {Zaanen}},\
  }\bibfield  {title} {\bibinfo {title} {{The space group classification of
  topological band-insulators}},\ }\href {https://doi.org/10.1038/nphys2513}
  {\bibfield  {journal} {\bibinfo  {journal} {Nat. Phys.}\ }\textbf {\bibinfo
  {volume} {9}},\ \bibinfo {pages} {98} (\bibinfo {year} {2012})}\BibitemShut
  {NoStop}%
\bibitem [{\citenamefont {Shiozaki}\ and\ \citenamefont
  {Sato}(2014)}]{Shiozaki2014}%
  \BibitemOpen
  \bibfield  {author} {\bibinfo {author} {\bibfnamefont {K.}~\bibnamefont
  {Shiozaki}}\ and\ \bibinfo {author} {\bibfnamefont {M.}~\bibnamefont
  {Sato}},\ }\bibfield  {title} {\bibinfo {title} {{Topology of crystalline
  insulators and superconductors}},\ }\href
  {https://doi.org/10.1103/PhysRevB.90.165114} {\bibfield  {journal} {\bibinfo
  {journal} {Phys. Rev. B}\ }\textbf {\bibinfo {volume} {90}},\ \bibinfo
  {pages} {165114} (\bibinfo {year} {2014})}\BibitemShut {NoStop}%
\bibitem [{\citenamefont {{Bradlyn}}\ \emph {et~al.}(2017)\citenamefont
  {{Bradlyn}}, \citenamefont {{Elcoro}}, \citenamefont {{Cano}}, \citenamefont
  {{Vergniory}}, \citenamefont {{Wang}}, \citenamefont {{Felser}},
  \citenamefont {{Aroyo}},\ and\ \citenamefont {{Bernevig}}}]{bernevig2017}%
  \BibitemOpen
  \bibfield  {author} {\bibinfo {author} {\bibfnamefont {B.}~\bibnamefont
  {{Bradlyn}}}, \bibinfo {author} {\bibfnamefont {L.}~\bibnamefont {{Elcoro}}},
  \bibinfo {author} {\bibfnamefont {J.}~\bibnamefont {{Cano}}}, \bibinfo
  {author} {\bibfnamefont {M.~G.}\ \bibnamefont {{Vergniory}}}, \bibinfo
  {author} {\bibfnamefont {Z.}~\bibnamefont {{Wang}}}, \bibinfo {author}
  {\bibfnamefont {C.}~\bibnamefont {{Felser}}}, \bibinfo {author}
  {\bibfnamefont {M.~I.}\ \bibnamefont {{Aroyo}}},\ and\ \bibinfo {author}
  {\bibfnamefont {B.~A.}\ \bibnamefont {{Bernevig}}},\ }\bibfield  {title}
  {\bibinfo {title} {{Topological quantum chemistry}},\ }\href
  {https://doi.org/10.1038/nature23268} {\bibfield  {journal} {\bibinfo
  {journal} {\nat}\ }\textbf {\bibinfo {volume} {547}},\ \bibinfo {pages} {298}
  (\bibinfo {year} {2017})}\BibitemShut {NoStop}%
\bibitem [{\citenamefont {Volovik}(2009)}]{Volovik2009}%
  \BibitemOpen
  \bibfield  {author} {\bibinfo {author} {\bibfnamefont {G.~E.}\ \bibnamefont
  {Volovik}},\ }\href
  {https://doi.org/10.1093/acprof:oso/9780199564842.001.0001} {\emph {\bibinfo
  {title} {{The Universe in a Helium Droplet}}}}\ (\bibinfo  {publisher}
  {Oxford University Press, Oxford, UK},\ \bibinfo {year} {2009})\BibitemShut
  {NoStop}%
\bibitem [{\citenamefont {{Ando}}\ and\ \citenamefont
  {{Fu}}(2015)}]{FuAndo2015}%
  \BibitemOpen
  \bibfield  {author} {\bibinfo {author} {\bibfnamefont {Y.}~\bibnamefont
  {{Ando}}}\ and\ \bibinfo {author} {\bibfnamefont {L.}~\bibnamefont {{Fu}}},\
  }\bibfield  {title} {\bibinfo {title} {{Topological Crystalline Insulators
  and Topological Superconductors: From Concepts to Materials}},\ }\href
  {https://doi.org/10.1146/annurev-conmatphys-031214-014501} {\bibfield
  {journal} {\bibinfo  {journal} {Annu. Rev. Condens. Matter Phys.}\ }\textbf
  {\bibinfo {volume} {6}},\ \bibinfo {pages} {361} (\bibinfo {year}
  {2015})}\BibitemShut {NoStop}%
\bibitem [{\citenamefont {Sato}\ and\ \citenamefont {Ando}(2017)}]{Sato2017}%
  \BibitemOpen
  \bibfield  {author} {\bibinfo {author} {\bibfnamefont {M.}~\bibnamefont
  {Sato}}\ and\ \bibinfo {author} {\bibfnamefont {Y.}~\bibnamefont {Ando}},\
  }\bibfield  {title} {\bibinfo {title} {{Topological superconductors: a
  review}},\ }\href {https://doi.org/10.1088/1361-6633/aa6ac7} {\bibfield
  {journal} {\bibinfo  {journal} {Rep. Prog. Phys.}\ }\textbf {\bibinfo
  {volume} {80}},\ \bibinfo {pages} {076501} (\bibinfo {year}
  {2017})}\BibitemShut {NoStop}%
\bibitem [{\citenamefont {Senthil}\ \emph {et~al.}(1999)\citenamefont
  {Senthil}, \citenamefont {Marston},\ and\ \citenamefont
  {Fisher}}]{senthilmarston1999}%
  \BibitemOpen
  \bibfield  {author} {\bibinfo {author} {\bibfnamefont {T.}~\bibnamefont
  {Senthil}}, \bibinfo {author} {\bibfnamefont {J.~B.}\ \bibnamefont
  {Marston}},\ and\ \bibinfo {author} {\bibfnamefont {M.~P.~A.}\ \bibnamefont
  {Fisher}},\ }\bibfield  {title} {\bibinfo {title} {{Spin quantum Hall effect
  in unconventional superconductors}},\ }\href
  {https://doi.org/10.1103/PhysRevB.60.4245} {\bibfield  {journal} {\bibinfo
  {journal} {Phys. Rev. B}\ }\textbf {\bibinfo {volume} {60}},\ \bibinfo
  {pages} {4245} (\bibinfo {year} {1999})}\BibitemShut {NoStop}%
\bibitem [{\citenamefont {Read}\ and\ \citenamefont {Green}(2000)}]{Read2000}%
  \BibitemOpen
  \bibfield  {author} {\bibinfo {author} {\bibfnamefont {N.}~\bibnamefont
  {Read}}\ and\ \bibinfo {author} {\bibfnamefont {D.}~\bibnamefont {Green}},\
  }\bibfield  {title} {\bibinfo {title} {{Paired states of fermions in two
  dimensions with breaking of parity and time-reversal symmetries and the
  fractional quantum {Hall} effect}},\ }\href
  {https://doi.org/10.1103/PhysRevB.61.10267} {\bibfield  {journal} {\bibinfo
  {journal} {Phys. Rev. B}\ }\textbf {\bibinfo {volume} {61}},\ \bibinfo
  {pages} {10267} (\bibinfo {year} {2000})}\BibitemShut {NoStop}%
\bibitem [{\citenamefont {Torres}(2019)}]{Torres2020}%
  \BibitemOpen
  \bibfield  {author} {\bibinfo {author} {\bibfnamefont {L.~E. F.~F.}\
  \bibnamefont {Torres}},\ }\bibfield  {title} {\bibinfo {title} {{Perspective
  on topological states of non-Hermitian lattices}},\ }\href
  {https://doi.org/10.1088/2515-7639/ab4092} {\bibfield  {journal} {\bibinfo
  {journal} {J. Phys. Mater.}\ }\textbf {\bibinfo {volume} {3}},\ \bibinfo
  {pages} {014002} (\bibinfo {year} {2019})}\BibitemShut {NoStop}%
\bibitem [{\citenamefont {Ghatak}\ and\ \citenamefont
  {Das}(2019)}]{Ghatak2019}%
  \BibitemOpen
  \bibfield  {author} {\bibinfo {author} {\bibfnamefont {A.}~\bibnamefont
  {Ghatak}}\ and\ \bibinfo {author} {\bibfnamefont {T.}~\bibnamefont {Das}},\
  }\bibfield  {title} {\bibinfo {title} {{New topological invariants in
  non-Hermitian systems}},\ }\href {https://doi.org/10.1088/1361-648X/ab11b3}
  {\bibfield  {journal} {\bibinfo  {journal} {J. Phys.: Condens. Matter}\
  }\textbf {\bibinfo {volume} {31}},\ \bibinfo {pages} {263001} (\bibinfo
  {year} {2019})}\BibitemShut {NoStop}%
\bibitem [{\citenamefont {Bergholtz}\ \emph {et~al.}(2021)\citenamefont
  {Bergholtz}, \citenamefont {Budich},\ and\ \citenamefont
  {Kunst}}]{Bergholtz2021}%
  \BibitemOpen
  \bibfield  {author} {\bibinfo {author} {\bibfnamefont {E.~J.}\ \bibnamefont
  {Bergholtz}}, \bibinfo {author} {\bibfnamefont {J.~C.}\ \bibnamefont
  {Budich}},\ and\ \bibinfo {author} {\bibfnamefont {F.~K.}\ \bibnamefont
  {Kunst}},\ }\bibfield  {title} {\bibinfo {title} {{Exceptional topology of
  non-Hermitian systems}},\ }\href
  {https://doi.org/10.1103/RevModPhys.93.015005} {\bibfield  {journal}
  {\bibinfo  {journal} {Rev. Mod. Phys.}\ }\textbf {\bibinfo {volume} {93}},\
  \bibinfo {pages} {015005} (\bibinfo {year} {2021})}\BibitemShut {NoStop}%
\bibitem [{\citenamefont {Esaki}\ \emph {et~al.}(2011)\citenamefont {Esaki},
  \citenamefont {Sato}, \citenamefont {Hasebe},\ and\ \citenamefont
  {Kohmoto}}]{Esaki2011}%
  \BibitemOpen
  \bibfield  {author} {\bibinfo {author} {\bibfnamefont {K.}~\bibnamefont
  {Esaki}}, \bibinfo {author} {\bibfnamefont {M.}~\bibnamefont {Sato}},
  \bibinfo {author} {\bibfnamefont {K.}~\bibnamefont {Hasebe}},\ and\ \bibinfo
  {author} {\bibfnamefont {M.}~\bibnamefont {Kohmoto}},\ }\bibfield  {title}
  {\bibinfo {title} {{Edge states and topological phases in non-Hermitian
  systems}},\ }\href {https://doi.org/10.1103/PhysRevB.84.205128} {\bibfield
  {journal} {\bibinfo  {journal} {Phys. Rev. B}\ }\textbf {\bibinfo {volume}
  {84}},\ \bibinfo {pages} {205128} (\bibinfo {year} {2011})}\BibitemShut
  {NoStop}%
\bibitem [{\citenamefont {Liang}\ and\ \citenamefont
  {Huang}(2013)}]{Liang2013}%
  \BibitemOpen
  \bibfield  {author} {\bibinfo {author} {\bibfnamefont {S.-D.}\ \bibnamefont
  {Liang}}\ and\ \bibinfo {author} {\bibfnamefont {G.-Y.}\ \bibnamefont
  {Huang}},\ }\bibfield  {title} {\bibinfo {title} {{Topological invariance and
  global Berry phase in non-Hermitian systems}},\ }\href
  {https://doi.org/10.1103/PhysRevA.87.012118} {\bibfield  {journal} {\bibinfo
  {journal} {Phys. Rev. A}\ }\textbf {\bibinfo {volume} {87}},\ \bibinfo
  {pages} {012118} (\bibinfo {year} {2013})}\BibitemShut {NoStop}%
\bibitem [{\citenamefont {Yao}\ and\ \citenamefont {Wang}(2018)}]{Yao2018Aug}%
  \BibitemOpen
  \bibfield  {author} {\bibinfo {author} {\bibfnamefont {S.}~\bibnamefont
  {Yao}}\ and\ \bibinfo {author} {\bibfnamefont {Z.}~\bibnamefont {Wang}},\
  }\bibfield  {title} {\bibinfo {title} {{Edge States and Topological
  Invariants of Non-Hermitian Systems}},\ }\href
  {https://doi.org/10.1103/PhysRevLett.121.086803} {\bibfield  {journal}
  {\bibinfo  {journal} {Phys. Rev. Lett.}\ }\textbf {\bibinfo {volume} {121}},\
  \bibinfo {pages} {086803} (\bibinfo {year} {2018})}\BibitemShut {NoStop}%
\bibitem [{\citenamefont {Yao}\ \emph {et~al.}(2018)\citenamefont {Yao},
  \citenamefont {Song},\ and\ \citenamefont {Wang}}]{Yao2018Sep}%
  \BibitemOpen
  \bibfield  {author} {\bibinfo {author} {\bibfnamefont {S.}~\bibnamefont
  {Yao}}, \bibinfo {author} {\bibfnamefont {F.}~\bibnamefont {Song}},\ and\
  \bibinfo {author} {\bibfnamefont {Z.}~\bibnamefont {Wang}},\ }\bibfield
  {title} {\bibinfo {title} {{Non-Hermitian Chern Bands}},\ }\href
  {https://doi.org/10.1103/PhysRevLett.121.136802} {\bibfield  {journal}
  {\bibinfo  {journal} {Phys. Rev. Lett.}\ }\textbf {\bibinfo {volume} {121}},\
  \bibinfo {pages} {136802} (\bibinfo {year} {2018})}\BibitemShut {NoStop}%
\bibitem [{\citenamefont {Gong}\ \emph {et~al.}(2018)\citenamefont {Gong},
  \citenamefont {Ashida}, \citenamefont {Kawabata}, \citenamefont {Takasan},
  \citenamefont {Higashikawa},\ and\ \citenamefont {Ueda}}]{Gong2018}%
  \BibitemOpen
  \bibfield  {author} {\bibinfo {author} {\bibfnamefont {Z.}~\bibnamefont
  {Gong}}, \bibinfo {author} {\bibfnamefont {Y.}~\bibnamefont {Ashida}},
  \bibinfo {author} {\bibfnamefont {K.}~\bibnamefont {Kawabata}}, \bibinfo
  {author} {\bibfnamefont {K.}~\bibnamefont {Takasan}}, \bibinfo {author}
  {\bibfnamefont {S.}~\bibnamefont {Higashikawa}},\ and\ \bibinfo {author}
  {\bibfnamefont {M.}~\bibnamefont {Ueda}},\ }\bibfield  {title} {\bibinfo
  {title} {{Topological Phases of Non-Hermitian Systems}},\ }\href
  {https://doi.org/10.1103/PhysRevX.8.031079} {\bibfield  {journal} {\bibinfo
  {journal} {Phys. Rev. X}\ }\textbf {\bibinfo {volume} {8}},\ \bibinfo {pages}
  {031079} (\bibinfo {year} {2018})}\BibitemShut {NoStop}%
\bibitem [{\citenamefont {Kawabata}\ \emph {et~al.}(2018)\citenamefont
  {Kawabata}, \citenamefont {Shiozaki},\ and\ \citenamefont
  {Ueda}}]{Kawabata2018}%
  \BibitemOpen
  \bibfield  {author} {\bibinfo {author} {\bibfnamefont {K.}~\bibnamefont
  {Kawabata}}, \bibinfo {author} {\bibfnamefont {K.}~\bibnamefont {Shiozaki}},\
  and\ \bibinfo {author} {\bibfnamefont {M.}~\bibnamefont {Ueda}},\ }\bibfield
  {title} {\bibinfo {title} {{Anomalous helical edge states in a non-Hermitian
  Chern insulator}},\ }\href {https://doi.org/10.1103/PhysRevB.98.165148}
  {\bibfield  {journal} {\bibinfo  {journal} {Phys. Rev. B}\ }\textbf {\bibinfo
  {volume} {98}},\ \bibinfo {pages} {165148} (\bibinfo {year}
  {2018})}\BibitemShut {NoStop}%
\bibitem [{\citenamefont {Shen}\ \emph {et~al.}(2018)\citenamefont {Shen},
  \citenamefont {Zhen},\ and\ \citenamefont {Fu}}]{Shen2018}%
  \BibitemOpen
  \bibfield  {author} {\bibinfo {author} {\bibfnamefont {H.}~\bibnamefont
  {Shen}}, \bibinfo {author} {\bibfnamefont {B.}~\bibnamefont {Zhen}},\ and\
  \bibinfo {author} {\bibfnamefont {L.}~\bibnamefont {Fu}},\ }\bibfield
  {title} {\bibinfo {title} {{Topological Band Theory for Non-Hermitian
  Hamiltonians}},\ }\href {https://doi.org/10.1103/PhysRevLett.120.146402}
  {\bibfield  {journal} {\bibinfo  {journal} {Phys. Rev. Lett.}\ }\textbf
  {\bibinfo {volume} {120}},\ \bibinfo {pages} {146402} (\bibinfo {year}
  {2018})}\BibitemShut {NoStop}%
\bibitem [{\citenamefont {Kunst}\ \emph {et~al.}(2018)\citenamefont {Kunst},
  \citenamefont {Edvardsson}, \citenamefont {Budich},\ and\ \citenamefont
  {Bergholtz}}]{Kunst2018}%
  \BibitemOpen
  \bibfield  {author} {\bibinfo {author} {\bibfnamefont {F.~K.}\ \bibnamefont
  {Kunst}}, \bibinfo {author} {\bibfnamefont {E.}~\bibnamefont {Edvardsson}},
  \bibinfo {author} {\bibfnamefont {J.~C.}\ \bibnamefont {Budich}},\ and\
  \bibinfo {author} {\bibfnamefont {E.~J.}\ \bibnamefont {Bergholtz}},\
  }\bibfield  {title} {\bibinfo {title} {{Biorthogonal Bulk-Boundary
  Correspondence in Non-Hermitian Systems}},\ }\href
  {https://doi.org/10.1103/PhysRevLett.121.026808} {\bibfield  {journal}
  {\bibinfo  {journal} {Phys. Rev. Lett.}\ }\textbf {\bibinfo {volume} {121}},\
  \bibinfo {pages} {026808} (\bibinfo {year} {2018})}\BibitemShut {NoStop}%
\bibitem [{\citenamefont {Kawabata}\ \emph
  {et~al.}(2019{\natexlab{a}})\citenamefont {Kawabata}, \citenamefont
  {Higashikawa}, \citenamefont {Gong}, \citenamefont {Ashida},\ and\
  \citenamefont {Ueda}}]{Kawabata2019}%
  \BibitemOpen
  \bibfield  {author} {\bibinfo {author} {\bibfnamefont {K.}~\bibnamefont
  {Kawabata}}, \bibinfo {author} {\bibfnamefont {S.}~\bibnamefont
  {Higashikawa}}, \bibinfo {author} {\bibfnamefont {Z.}~\bibnamefont {Gong}},
  \bibinfo {author} {\bibfnamefont {Y.}~\bibnamefont {Ashida}},\ and\ \bibinfo
  {author} {\bibfnamefont {M.}~\bibnamefont {Ueda}},\ }\bibfield  {title}
  {\bibinfo {title} {{Topological unification of time-reversal and
  particle-hole symmetries in non-Hermitian physics}},\ }\href
  {https://doi.org/10.1038/s41467-018-08254-y} {\bibfield  {journal} {\bibinfo
  {journal} {Nat. Commun.}\ }\textbf {\bibinfo {volume} {10}},\ \bibinfo
  {pages} {297} (\bibinfo {year} {2019}{\natexlab{a}})}\BibitemShut {NoStop}%
\bibitem [{\citenamefont {Zhou}\ and\ \citenamefont {Lee}(2019)}]{Zhou2019}%
  \BibitemOpen
  \bibfield  {author} {\bibinfo {author} {\bibfnamefont {H.}~\bibnamefont
  {Zhou}}\ and\ \bibinfo {author} {\bibfnamefont {J.~Y.}\ \bibnamefont {Lee}},\
  }\bibfield  {title} {\bibinfo {title} {{Periodic table for topological bands
  with non-Hermitian symmetries}},\ }\href
  {https://doi.org/10.1103/PhysRevB.99.235112} {\bibfield  {journal} {\bibinfo
  {journal} {Phys. Rev. B}\ }\textbf {\bibinfo {volume} {99}},\ \bibinfo
  {pages} {235112} (\bibinfo {year} {2019})}\BibitemShut {NoStop}%
\bibitem [{\citenamefont {Yokomizo}\ and\ \citenamefont
  {Murakami}(2019)}]{Yokomizo2019}%
  \BibitemOpen
  \bibfield  {author} {\bibinfo {author} {\bibfnamefont {K.}~\bibnamefont
  {Yokomizo}}\ and\ \bibinfo {author} {\bibfnamefont {S.}~\bibnamefont
  {Murakami}},\ }\bibfield  {title} {\bibinfo {title} {{Non-Bloch Band Theory
  of Non-Hermitian Systems}},\ }\href
  {https://doi.org/10.1103/PhysRevLett.123.066404} {\bibfield  {journal}
  {\bibinfo  {journal} {Phys. Rev. Lett.}\ }\textbf {\bibinfo {volume} {123}},\
  \bibinfo {pages} {066404} (\bibinfo {year} {2019})}\BibitemShut {NoStop}%
\bibitem [{\citenamefont {Kawabata}\ \emph
  {et~al.}(2019{\natexlab{b}})\citenamefont {Kawabata}, \citenamefont
  {Shiozaki}, \citenamefont {Ueda},\ and\ \citenamefont
  {Sato}}]{kawabata2019b}%
  \BibitemOpen
  \bibfield  {author} {\bibinfo {author} {\bibfnamefont {K.}~\bibnamefont
  {Kawabata}}, \bibinfo {author} {\bibfnamefont {K.}~\bibnamefont {Shiozaki}},
  \bibinfo {author} {\bibfnamefont {M.}~\bibnamefont {Ueda}},\ and\ \bibinfo
  {author} {\bibfnamefont {M.}~\bibnamefont {Sato}},\ }\bibfield  {title}
  {\bibinfo {title} {Symmetry and topology in non-hermitian physics},\ }\href
  {https://doi.org/10.1103/PhysRevX.9.041015} {\bibfield  {journal} {\bibinfo
  {journal} {Phys. Rev. X}\ }\textbf {\bibinfo {volume} {9}},\ \bibinfo {pages}
  {041015} (\bibinfo {year} {2019}{\natexlab{b}})}\BibitemShut {NoStop}%
\bibitem [{\citenamefont {Lee}\ and\ \citenamefont {Thomale}(2019)}]{Lee2019}%
  \BibitemOpen
  \bibfield  {author} {\bibinfo {author} {\bibfnamefont {C.~H.}\ \bibnamefont
  {Lee}}\ and\ \bibinfo {author} {\bibfnamefont {R.}~\bibnamefont {Thomale}},\
  }\bibfield  {title} {\bibinfo {title} {{Anatomy of skin modes and topology in
  non-Hermitian systems}},\ }\href {https://doi.org/10.1103/PhysRevB.99.201103}
  {\bibfield  {journal} {\bibinfo  {journal} {Phys. Rev. B}\ }\textbf {\bibinfo
  {volume} {99}},\ \bibinfo {pages} {201103} (\bibinfo {year}
  {2019})}\BibitemShut {NoStop}%
\bibitem [{\citenamefont {Okuma}\ \emph {et~al.}(2020)\citenamefont {Okuma},
  \citenamefont {Kawabata}, \citenamefont {Shiozaki},\ and\ \citenamefont
  {Sato}}]{Okuma2020}%
  \BibitemOpen
  \bibfield  {author} {\bibinfo {author} {\bibfnamefont {N.}~\bibnamefont
  {Okuma}}, \bibinfo {author} {\bibfnamefont {K.}~\bibnamefont {Kawabata}},
  \bibinfo {author} {\bibfnamefont {K.}~\bibnamefont {Shiozaki}},\ and\
  \bibinfo {author} {\bibfnamefont {M.}~\bibnamefont {Sato}},\ }\bibfield
  {title} {\bibinfo {title} {{Topological Origin of Non-Hermitian Skin
  Effects}},\ }\href {https://doi.org/10.1103/PhysRevLett.124.086801}
  {\bibfield  {journal} {\bibinfo  {journal} {Phys. Rev. Lett.}\ }\textbf
  {\bibinfo {volume} {124}},\ \bibinfo {pages} {086801} (\bibinfo {year}
  {2020})}\BibitemShut {NoStop}%
\bibitem [{\citenamefont {Zhang}\ \emph {et~al.}(2020)\citenamefont {Zhang},
  \citenamefont {Yang},\ and\ \citenamefont {Fang}}]{Zhang2020Sep}%
  \BibitemOpen
  \bibfield  {author} {\bibinfo {author} {\bibfnamefont {K.}~\bibnamefont
  {Zhang}}, \bibinfo {author} {\bibfnamefont {Z.}~\bibnamefont {Yang}},\ and\
  \bibinfo {author} {\bibfnamefont {C.}~\bibnamefont {Fang}},\ }\bibfield
  {title} {\bibinfo {title} {{Correspondence between Winding Numbers and Skin
  Modes in Non-Hermitian Systems}},\ }\href
  {https://doi.org/10.1103/PhysRevLett.125.126402} {\bibfield  {journal}
  {\bibinfo  {journal} {Phys. Rev. Lett.}\ }\textbf {\bibinfo {volume} {125}},\
  \bibinfo {pages} {126402} (\bibinfo {year} {2020})}\BibitemShut {NoStop}%
\bibitem [{\citenamefont {Borgnia}\ \emph {et~al.}(2020)\citenamefont
  {Borgnia}, \citenamefont {Kruchkov},\ and\ \citenamefont
  {Slager}}]{Borgnia2020}%
  \BibitemOpen
  \bibfield  {author} {\bibinfo {author} {\bibfnamefont {D.~S.}\ \bibnamefont
  {Borgnia}}, \bibinfo {author} {\bibfnamefont {A.~J.}\ \bibnamefont
  {Kruchkov}},\ and\ \bibinfo {author} {\bibfnamefont {R.-J.}\ \bibnamefont
  {Slager}},\ }\bibfield  {title} {\bibinfo {title} {{Non-Hermitian Boundary
  Modes and Topology}},\ }\href
  {https://doi.org/10.1103/PhysRevLett.124.056802} {\bibfield  {journal}
  {\bibinfo  {journal} {Phys. Rev. Lett.}\ }\textbf {\bibinfo {volume} {124}},\
  \bibinfo {pages} {056802} (\bibinfo {year} {2020})}\BibitemShut {NoStop}%
\bibitem [{\citenamefont {Wu}\ \emph {et~al.}(2020)\citenamefont {Wu},
  \citenamefont {Jin},\ and\ \citenamefont {Song}}]{Wu2020}%
  \BibitemOpen
  \bibfield  {author} {\bibinfo {author} {\bibfnamefont {H.~C.}\ \bibnamefont
  {Wu}}, \bibinfo {author} {\bibfnamefont {L.}~\bibnamefont {Jin}},\ and\
  \bibinfo {author} {\bibfnamefont {Z.}~\bibnamefont {Song}},\ }\bibfield
  {title} {\bibinfo {title} {{Nontrivial topological phase with a zero Chern
  number}},\ }\href {https://doi.org/10.1103/PhysRevB.102.035145} {\bibfield
  {journal} {\bibinfo  {journal} {Phys. Rev. B}\ }\textbf {\bibinfo {volume}
  {102}},\ \bibinfo {pages} {035145} (\bibinfo {year} {2020})}\BibitemShut
  {NoStop}%
\bibitem [{\citenamefont {Sun}\ \emph {et~al.}(2021)\citenamefont {Sun},
  \citenamefont {Zhu},\ and\ \citenamefont {Hughes}}]{Sun2021}%
  \BibitemOpen
  \bibfield  {author} {\bibinfo {author} {\bibfnamefont {X.-Q.}\ \bibnamefont
  {Sun}}, \bibinfo {author} {\bibfnamefont {P.}~\bibnamefont {Zhu}},\ and\
  \bibinfo {author} {\bibfnamefont {T.~L.}\ \bibnamefont {Hughes}},\ }\bibfield
   {title} {\bibinfo {title} {{Geometric Response and Disclination-Induced Skin
  Effects in Non-Hermitian Systems}},\ }\href
  {https://doi.org/10.1103/PhysRevLett.127.066401} {\bibfield  {journal}
  {\bibinfo  {journal} {Phys. Rev. Lett.}\ }\textbf {\bibinfo {volume} {127}},\
  \bibinfo {pages} {066401} (\bibinfo {year} {2021})}\BibitemShut {NoStop}%
\bibitem [{\citenamefont {Zhu}\ \emph {et~al.}(2021)\citenamefont {Zhu},
  \citenamefont {Teo}, \citenamefont {Li},\ and\ \citenamefont
  {Gong}}]{Zhu2021}%
  \BibitemOpen
  \bibfield  {author} {\bibinfo {author} {\bibfnamefont {W.}~\bibnamefont
  {Zhu}}, \bibinfo {author} {\bibfnamefont {W.~X.}\ \bibnamefont {Teo}},
  \bibinfo {author} {\bibfnamefont {L.}~\bibnamefont {Li}},\ and\ \bibinfo
  {author} {\bibfnamefont {J.}~\bibnamefont {Gong}},\ }\bibfield  {title}
  {\bibinfo {title} {{Delocalization of topological edge states}},\ }\href
  {https://doi.org/10.1103/PhysRevB.103.195414} {\bibfield  {journal} {\bibinfo
   {journal} {Phys. Rev. B}\ }\textbf {\bibinfo {volume} {103}},\ \bibinfo
  {pages} {195414} (\bibinfo {year} {2021})}\BibitemShut {NoStop}%
\bibitem [{\citenamefont {Panigrahi}\ \emph {et~al.}(2022)\citenamefont
  {Panigrahi}, \citenamefont {Moessner},\ and\ \citenamefont
  {Roy}}]{PanigraphiNH2022}%
  \BibitemOpen
  \bibfield  {author} {\bibinfo {author} {\bibfnamefont {A.}~\bibnamefont
  {Panigrahi}}, \bibinfo {author} {\bibfnamefont {R.}~\bibnamefont
  {Moessner}},\ and\ \bibinfo {author} {\bibfnamefont {B.}~\bibnamefont
  {Roy}},\ }\bibfield  {title} {\bibinfo {title} {{Non-Hermitian dislocation
  modes: Stability and melting across exceptional points}},\ }\href
  {https://doi.org/10.1103/PhysRevB.106.L041302} {\bibfield  {journal}
  {\bibinfo  {journal} {Phys. Rev. B}\ }\textbf {\bibinfo {volume} {106}},\
  \bibinfo {pages} {L041302} (\bibinfo {year} {2022})}\BibitemShut {NoStop}%
\bibitem [{\citenamefont {Manna}\ and\ \citenamefont {Roy}(2023)}]{Manna2023}%
  \BibitemOpen
  \bibfield  {author} {\bibinfo {author} {\bibfnamefont {S.}~\bibnamefont
  {Manna}}\ and\ \bibinfo {author} {\bibfnamefont {B.}~\bibnamefont {Roy}},\
  }\bibfield  {title} {\bibinfo {title} {{Inner skin effects on non-Hermitian
  topological fractals}},\ }\href {https://doi.org/10.1038/s42005-023-01130-2}
  {\bibfield  {journal} {\bibinfo  {journal} {Commun. Phys.}\ }\textbf
  {\bibinfo {volume} {6}},\ \bibinfo {pages} {10} (\bibinfo {year}
  {2023})}\BibitemShut {NoStop}%
\bibitem [{\citenamefont {{Chaturvedi}}\ \emph {et~al.}()\citenamefont
  {{Chaturvedi}}, \citenamefont {{K{\"o}nye}}, \citenamefont {{Hankiewicz}},
  \citenamefont {{van den Brink}},\ and\ \citenamefont {{Cosma
  Fulga}}}]{CosmaNHtrans2024}%
  \BibitemOpen
  \bibfield  {author} {\bibinfo {author} {\bibfnamefont {R.}~\bibnamefont
  {{Chaturvedi}}}, \bibinfo {author} {\bibfnamefont {V.}~\bibnamefont
  {{K{\"o}nye}}}, \bibinfo {author} {\bibfnamefont {E.~M.}\ \bibnamefont
  {{Hankiewicz}}}, \bibinfo {author} {\bibfnamefont {J.}~\bibnamefont {{van den
  Brink}}},\ and\ \bibinfo {author} {\bibfnamefont {I.}~\bibnamefont {{Cosma
  Fulga}}},\ }\href@noop {} {\bibinfo {title} {{Non-Hermitian topology of
  transport in Chern insulators}}},\ \Eprint
  {https://arxiv.org/abs/arXiv:2406.14303} {arXiv:2406.14303} \BibitemShut
  {NoStop}%
\bibitem [{\citenamefont {Klitzing}\ \emph {et~al.}(1980)\citenamefont
  {Klitzing}, \citenamefont {Dorda},\ and\ \citenamefont
  {Pepper}}]{Klitzing1980}%
  \BibitemOpen
  \bibfield  {author} {\bibinfo {author} {\bibfnamefont {K.~v.}\ \bibnamefont
  {Klitzing}}, \bibinfo {author} {\bibfnamefont {G.}~\bibnamefont {Dorda}},\
  and\ \bibinfo {author} {\bibfnamefont {M.}~\bibnamefont {Pepper}},\
  }\bibfield  {title} {\bibinfo {title} {{New Method for High-Accuracy
  Determination of the Fine-Structure Constant Based on Quantized Hall
  Resistance}},\ }\href {https://doi.org/10.1103/PhysRevLett.45.494} {\bibfield
   {journal} {\bibinfo  {journal} {Phys. Rev. Lett.}\ }\textbf {\bibinfo
  {volume} {45}},\ \bibinfo {pages} {494} (\bibinfo {year} {1980})}\BibitemShut
  {NoStop}%
\bibitem [{\citenamefont {Ko\"onig}\ \emph {et~al.}(2007)\citenamefont
  {Ko\"onig}, \citenamefont {Wiedmann}, \citenamefont {Br\"une}, \citenamefont
  {Roth}, \citenamefont {Buhmann}, \citenamefont {Molenkamp}, \citenamefont
  {Qi},\ and\ \citenamefont {Zhang}}]{Konig2007}%
  \BibitemOpen
  \bibfield  {author} {\bibinfo {author} {\bibfnamefont {M.}~\bibnamefont
  {Ko\"onig}}, \bibinfo {author} {\bibfnamefont {S.}~\bibnamefont {Wiedmann}},
  \bibinfo {author} {\bibfnamefont {C.}~\bibnamefont {Br\"une}}, \bibinfo
  {author} {\bibfnamefont {A.}~\bibnamefont {Roth}}, \bibinfo {author}
  {\bibfnamefont {H.}~\bibnamefont {Buhmann}}, \bibinfo {author} {\bibfnamefont
  {L.~W.}\ \bibnamefont {Molenkamp}}, \bibinfo {author} {\bibfnamefont {X.-L.}\
  \bibnamefont {Qi}},\ and\ \bibinfo {author} {\bibfnamefont {S.-C.}\
  \bibnamefont {Zhang}},\ }\bibfield  {title} {\bibinfo {title} {{Quantum Spin
  Hall Insulator State in {HgTe} Quantum Wells}},\ }\href
  {https://doi.org/10.1126/science.1148047} {\bibfield  {journal} {\bibinfo
  {journal} {Science}\ }\textbf {\bibinfo {volume} {318}},\ \bibinfo {pages}
  {766} (\bibinfo {year} {2007})}\BibitemShut {NoStop}%
\bibitem [{\citenamefont {Knez}\ \emph {et~al.}(2011)\citenamefont {Knez},
  \citenamefont {Du},\ and\ \citenamefont {Sullivan}}]{Knez2011}%
  \BibitemOpen
  \bibfield  {author} {\bibinfo {author} {\bibfnamefont {I.}~\bibnamefont
  {Knez}}, \bibinfo {author} {\bibfnamefont {R.-R.}\ \bibnamefont {Du}},\ and\
  \bibinfo {author} {\bibfnamefont {G.}~\bibnamefont {Sullivan}},\ }\bibfield
  {title} {\bibinfo {title} {{Evidence for Helical Edge Modes in Inverted
  $\mathrm{InAs}/\mathrm{GaSb}$ Quantum Wells}},\ }\href
  {https://doi.org/10.1103/PhysRevLett.107.136603} {\bibfield  {journal}
  {\bibinfo  {journal} {Phys. Rev. Lett.}\ }\textbf {\bibinfo {volume} {107}},\
  \bibinfo {pages} {136603} (\bibinfo {year} {2011})}\BibitemShut {NoStop}%
\bibitem [{\citenamefont {{Yu}}\ \emph {et~al.}(2010)\citenamefont {{Yu}},
  \citenamefont {{Zhang}}, \citenamefont {{Zhang}}, \citenamefont {{Zhang}},
  \citenamefont {{Dai}},\ and\ \citenamefont {{Fang}}}]{ZhongFang2010}%
  \BibitemOpen
  \bibfield  {author} {\bibinfo {author} {\bibfnamefont {R.}~\bibnamefont
  {{Yu}}}, \bibinfo {author} {\bibfnamefont {W.}~\bibnamefont {{Zhang}}},
  \bibinfo {author} {\bibfnamefont {H.-J.}\ \bibnamefont {{Zhang}}}, \bibinfo
  {author} {\bibfnamefont {S.-C.}\ \bibnamefont {{Zhang}}}, \bibinfo {author}
  {\bibfnamefont {X.}~\bibnamefont {{Dai}}},\ and\ \bibinfo {author}
  {\bibfnamefont {Z.}~\bibnamefont {{Fang}}},\ }\bibfield  {title} {\bibinfo
  {title} {{Quantized Anomalous Hall Effect in Magnetic Topological
  Insulators}},\ }\href {https://doi.org/10.1126/science.1187485} {\bibfield
  {journal} {\bibinfo  {journal} {Science}\ }\textbf {\bibinfo {volume}
  {329}},\ \bibinfo {pages} {61} (\bibinfo {year} {2010})}\BibitemShut
  {NoStop}%
\bibitem [{\citenamefont {{Chang}}\ \emph {et~al.}(2013)\citenamefont
  {{Chang}}, \citenamefont {{Zhang}}, \citenamefont {{Feng}}, \citenamefont
  {{Shen}}, \citenamefont {{Zhang}}, \citenamefont {{Guo}}, \citenamefont
  {{Li}}, \citenamefont {{Ou}}, \citenamefont {{Wei}}, \citenamefont {{Wang}},
  \citenamefont {{Ji}}, \citenamefont {{Feng}}, \citenamefont {{Ji}},
  \citenamefont {{Chen}}, \citenamefont {{Jia}}, \citenamefont {{Dai}},
  \citenamefont {{Fang}}, \citenamefont {{Zhang}}, \citenamefont {{He}},
  \citenamefont {{Wang}}, \citenamefont {{Lu}}, \citenamefont {{Ma}},\ and\
  \citenamefont {{Xue}}}]{QKXu2013}%
  \BibitemOpen
  \bibfield  {author} {\bibinfo {author} {\bibfnamefont {C.-Z.}\ \bibnamefont
  {{Chang}}}, \bibinfo {author} {\bibfnamefont {J.}~\bibnamefont {{Zhang}}},
  \bibinfo {author} {\bibfnamefont {X.}~\bibnamefont {{Feng}}}, \bibinfo
  {author} {\bibfnamefont {J.}~\bibnamefont {{Shen}}}, \bibinfo {author}
  {\bibfnamefont {Z.}~\bibnamefont {{Zhang}}}, \bibinfo {author} {\bibfnamefont
  {M.}~\bibnamefont {{Guo}}}, \bibinfo {author} {\bibfnamefont
  {K.}~\bibnamefont {{Li}}}, \bibinfo {author} {\bibfnamefont {Y.}~\bibnamefont
  {{Ou}}}, \bibinfo {author} {\bibfnamefont {P.}~\bibnamefont {{Wei}}},
  \bibinfo {author} {\bibfnamefont {L.-L.}\ \bibnamefont {{Wang}}}, \bibinfo
  {author} {\bibfnamefont {Z.-Q.}\ \bibnamefont {{Ji}}}, \bibinfo {author}
  {\bibfnamefont {Y.}~\bibnamefont {{Feng}}}, \bibinfo {author} {\bibfnamefont
  {S.}~\bibnamefont {{Ji}}}, \bibinfo {author} {\bibfnamefont {X.}~\bibnamefont
  {{Chen}}}, \bibinfo {author} {\bibfnamefont {J.}~\bibnamefont {{Jia}}},
  \bibinfo {author} {\bibfnamefont {X.}~\bibnamefont {{Dai}}}, \bibinfo
  {author} {\bibfnamefont {Z.}~\bibnamefont {{Fang}}}, \bibinfo {author}
  {\bibfnamefont {S.-C.}\ \bibnamefont {{Zhang}}}, \bibinfo {author}
  {\bibfnamefont {K.}~\bibnamefont {{He}}}, \bibinfo {author} {\bibfnamefont
  {Y.}~\bibnamefont {{Wang}}}, \bibinfo {author} {\bibfnamefont
  {L.}~\bibnamefont {{Lu}}}, \bibinfo {author} {\bibfnamefont {X.-C.}\
  \bibnamefont {{Ma}}},\ and\ \bibinfo {author} {\bibfnamefont {Q.-K.}\
  \bibnamefont {{Xue}}},\ }\bibfield  {title} {\bibinfo {title} {{Experimental
  Observation of the Quantum Anomalous Hall Effect in a Magnetic Topological
  Insulator}},\ }\href {https://doi.org/10.1126/science.1234414} {\bibfield
  {journal} {\bibinfo  {journal} {Science}\ }\textbf {\bibinfo {volume}
  {340}},\ \bibinfo {pages} {167} (\bibinfo {year} {2013})}\BibitemShut
  {NoStop}%
\bibitem [{\citenamefont {Chang}\ \emph {et~al.}(2015)\citenamefont {Chang},
  \citenamefont {Zhao}, \citenamefont {Kim}, \citenamefont {Zhang},
  \citenamefont {Assaf}, \citenamefont {Heiman}, \citenamefont {Zhang},
  \citenamefont {Liu}, \citenamefont {Chan},\ and\ \citenamefont
  {Moodera}}]{Chang2015}%
  \BibitemOpen
  \bibfield  {author} {\bibinfo {author} {\bibfnamefont {C.-Z.}\ \bibnamefont
  {Chang}}, \bibinfo {author} {\bibfnamefont {W.}~\bibnamefont {Zhao}},
  \bibinfo {author} {\bibfnamefont {D.~Y.}\ \bibnamefont {Kim}}, \bibinfo
  {author} {\bibfnamefont {H.}~\bibnamefont {Zhang}}, \bibinfo {author}
  {\bibfnamefont {B.~A.}\ \bibnamefont {Assaf}}, \bibinfo {author}
  {\bibfnamefont {D.}~\bibnamefont {Heiman}}, \bibinfo {author} {\bibfnamefont
  {S.-C.}\ \bibnamefont {Zhang}}, \bibinfo {author} {\bibfnamefont
  {C.}~\bibnamefont {Liu}}, \bibinfo {author} {\bibfnamefont {M.~H.~W.}\
  \bibnamefont {Chan}},\ and\ \bibinfo {author} {\bibfnamefont {J.~S.}\
  \bibnamefont {Moodera}},\ }\bibfield  {title} {\bibinfo {title}
  {{High-precision realization of robust quantum anomalous Hall state in a hard
  ferromagnetic topological insulator}},\ }\href
  {https://doi.org/10.1038/nmat4204} {\bibfield  {journal} {\bibinfo  {journal}
  {Nature Mater}\ }\textbf {\bibinfo {volume} {14}},\ \bibinfo {pages} {473}
  (\bibinfo {year} {2015})}\BibitemShut {NoStop}%
\bibitem [{\citenamefont {{Banerjee}}\ \emph {et~al.}(2018)\citenamefont
  {{Banerjee}}, \citenamefont {{Heiblum}}, \citenamefont {{Umansky}},
  \citenamefont {{Feldman}}, \citenamefont {{Oreg}},\ and\ \citenamefont
  {{Stern}}}]{Banerjee2018}%
  \BibitemOpen
  \bibfield  {author} {\bibinfo {author} {\bibfnamefont {M.}~\bibnamefont
  {{Banerjee}}}, \bibinfo {author} {\bibfnamefont {M.}~\bibnamefont
  {{Heiblum}}}, \bibinfo {author} {\bibfnamefont {V.}~\bibnamefont
  {{Umansky}}}, \bibinfo {author} {\bibfnamefont {D.~E.}\ \bibnamefont
  {{Feldman}}}, \bibinfo {author} {\bibfnamefont {Y.}~\bibnamefont {{Oreg}}},\
  and\ \bibinfo {author} {\bibfnamefont {A.}~\bibnamefont {{Stern}}},\
  }\bibfield  {title} {\bibinfo {title} {{Observation of half-integer thermal
  Hall conductance}},\ }\href {https://doi.org/10.1038/s41586-018-0184-1}
  {\bibfield  {journal} {\bibinfo  {journal} {\nat}\ }\textbf {\bibinfo
  {volume} {559}},\ \bibinfo {pages} {205} (\bibinfo {year}
  {2018})}\BibitemShut {NoStop}%
\bibitem [{\citenamefont {Kasahara}\ \emph {et~al.}(2018)\citenamefont
  {Kasahara}, \citenamefont {Ohnishi}, \citenamefont {Mizukami}, \citenamefont
  {Tanaka}, \citenamefont {Ma}, \citenamefont {Sugii}, \citenamefont {Kurita},
  \citenamefont {Tanaka}, \citenamefont {Nasu}, \citenamefont {Motome},
  \citenamefont {Shibauchi},\ and\ \citenamefont {Matsuda}}]{Kasahara2018}%
  \BibitemOpen
  \bibfield  {author} {\bibinfo {author} {\bibfnamefont {Y.}~\bibnamefont
  {Kasahara}}, \bibinfo {author} {\bibfnamefont {T.}~\bibnamefont {Ohnishi}},
  \bibinfo {author} {\bibfnamefont {Y.}~\bibnamefont {Mizukami}}, \bibinfo
  {author} {\bibfnamefont {O.}~\bibnamefont {Tanaka}}, \bibinfo {author}
  {\bibfnamefont {S.}~\bibnamefont {Ma}}, \bibinfo {author} {\bibfnamefont
  {K.}~\bibnamefont {Sugii}}, \bibinfo {author} {\bibfnamefont
  {N.}~\bibnamefont {Kurita}}, \bibinfo {author} {\bibfnamefont
  {H.}~\bibnamefont {Tanaka}}, \bibinfo {author} {\bibfnamefont
  {J.}~\bibnamefont {Nasu}}, \bibinfo {author} {\bibfnamefont {Y.}~\bibnamefont
  {Motome}}, \bibinfo {author} {\bibfnamefont {T.}~\bibnamefont {Shibauchi}},\
  and\ \bibinfo {author} {\bibfnamefont {Y.}~\bibnamefont {Matsuda}},\
  }\bibfield  {title} {\bibinfo {title} {{Majorana quantization and
  half-integer thermal quantum Hall effect in a Kitaev spin liquid}},\ }\href
  {https://doi.org/10.1038/s41586-018-0274-0} {\bibfield  {journal} {\bibinfo
  {journal} {\nat}\ }\textbf {\bibinfo {volume} {559}},\ \bibinfo {pages} {227}
  (\bibinfo {year} {2018})}\BibitemShut {NoStop}%
\bibitem [{\citenamefont {{Srivastav}}\ \emph {et~al.}(2019)\citenamefont
  {{Srivastav}}, \citenamefont {{Sahu}}, \citenamefont {{Watanabe}},
  \citenamefont {{Taniguchi}}, \citenamefont {{Banerjee}},\ and\ \citenamefont
  {{Das}}}]{Srivastav2019}%
  \BibitemOpen
  \bibfield  {author} {\bibinfo {author} {\bibfnamefont {S.~K.}\ \bibnamefont
  {{Srivastav}}}, \bibinfo {author} {\bibfnamefont {M.~R.}\ \bibnamefont
  {{Sahu}}}, \bibinfo {author} {\bibfnamefont {K.}~\bibnamefont {{Watanabe}}},
  \bibinfo {author} {\bibfnamefont {T.}~\bibnamefont {{Taniguchi}}}, \bibinfo
  {author} {\bibfnamefont {S.}~\bibnamefont {{Banerjee}}},\ and\ \bibinfo
  {author} {\bibfnamefont {A.}~\bibnamefont {{Das}}},\ }\bibfield  {title}
  {\bibinfo {title} {{Universal quantized thermal conductance in graphene}},\
  }\href {https://doi.org/10.1126/sciadv.aaw5798} {\bibfield  {journal}
  {\bibinfo  {journal} {Sci. Adv.}\ }\textbf {\bibinfo {volume} {5}},\ \bibinfo
  {eid} {eaaw5798} (\bibinfo {year} {2019})}\BibitemShut {NoStop}%
\bibitem [{\citenamefont {Le~Breton}\ \emph {et~al.}(2022)\citenamefont
  {Le~Breton}, \citenamefont {Delagrange}, \citenamefont {Hong}, \citenamefont
  {Garg}, \citenamefont {Watanabe}, \citenamefont {Taniguchi}, \citenamefont
  {Ribeiro-Palau}, \citenamefont {Roulleau}, \citenamefont {Roche},\ and\
  \citenamefont {Parmentier}}]{Breton2022}%
  \BibitemOpen
  \bibfield  {author} {\bibinfo {author} {\bibfnamefont {G.}~\bibnamefont
  {Le~Breton}}, \bibinfo {author} {\bibfnamefont {R.}~\bibnamefont
  {Delagrange}}, \bibinfo {author} {\bibfnamefont {Y.}~\bibnamefont {Hong}},
  \bibinfo {author} {\bibfnamefont {M.}~\bibnamefont {Garg}}, \bibinfo {author}
  {\bibfnamefont {K.}~\bibnamefont {Watanabe}}, \bibinfo {author}
  {\bibfnamefont {T.}~\bibnamefont {Taniguchi}}, \bibinfo {author}
  {\bibfnamefont {R.}~\bibnamefont {Ribeiro-Palau}}, \bibinfo {author}
  {\bibfnamefont {P.}~\bibnamefont {Roulleau}}, \bibinfo {author}
  {\bibfnamefont {P.}~\bibnamefont {Roche}},\ and\ \bibinfo {author}
  {\bibfnamefont {F.~D.}\ \bibnamefont {Parmentier}},\ }\bibfield  {title}
  {\bibinfo {title} {{Heat Equilibration of Integer and Fractional Quantum Hall
  Edge Modes in Graphene}},\ }\href
  {https://doi.org/10.1103/PhysRevLett.129.116803} {\bibfield  {journal}
  {\bibinfo  {journal} {Phys. Rev. Lett.}\ }\textbf {\bibinfo {volume} {129}},\
  \bibinfo {pages} {116803} (\bibinfo {year} {2022})}\BibitemShut {NoStop}%
\bibitem [{\citenamefont {Altland}\ and\ \citenamefont
  {Zirnbauer}(1997)}]{Altland1997}%
  \BibitemOpen
  \bibfield  {author} {\bibinfo {author} {\bibfnamefont {A.}~\bibnamefont
  {Altland}}\ and\ \bibinfo {author} {\bibfnamefont {M.~R.}\ \bibnamefont
  {Zirnbauer}},\ }\bibfield  {title} {\bibinfo {title} {{Nonstandard symmetry
  classes in mesoscopic normal-superconducting hybrid structures}},\ }\href
  {https://doi.org/10.1103/PhysRevB.55.1142} {\bibfield  {journal} {\bibinfo
  {journal} {Phys. Rev. B}\ }\textbf {\bibinfo {volume} {55}},\ \bibinfo
  {pages} {1142} (\bibinfo {year} {1997})}\BibitemShut {NoStop}%
\bibitem [{\citenamefont {Salib}\ \emph {et~al.}()\citenamefont {Salib},
  \citenamefont {Das},\ and\ \citenamefont {Roy}}]{SalibDasRoy2023}%
  \BibitemOpen
  \bibfield  {author} {\bibinfo {author} {\bibfnamefont {D.~J.}\ \bibnamefont
  {Salib}}, \bibinfo {author} {\bibfnamefont {S.~K.}\ \bibnamefont {Das}},\
  and\ \bibinfo {author} {\bibfnamefont {B.}~\bibnamefont {Roy}},\ }\href@noop
  {} {\bibinfo {title} {{Model non-Hermitian topological operators without skin
  effect}}},\ \Eprint {https://arxiv.org/abs/arXiv:2309.12310}
  {arXiv:2309.12310} \BibitemShut {NoStop}%
\bibitem [{\citenamefont {Groth}\ \emph {et~al.}(2014)\citenamefont {Groth},
  \citenamefont {Wimmer}, \citenamefont {Akhmerov},\ and\ \citenamefont
  {Waintal}}]{Groth2014}%
  \BibitemOpen
  \bibfield  {author} {\bibinfo {author} {\bibfnamefont {C.~W.}\ \bibnamefont
  {Groth}}, \bibinfo {author} {\bibfnamefont {M.}~\bibnamefont {Wimmer}},
  \bibinfo {author} {\bibfnamefont {A.~R.}\ \bibnamefont {Akhmerov}},\ and\
  \bibinfo {author} {\bibfnamefont {X.}~\bibnamefont {Waintal}},\ }\bibfield
  {title} {\bibinfo {title} {{Kwant: A software package for quantum
  transport}},\ }\href {https://doi.org/10.1088/1367-2630/16/6/063065}
  {\bibfield  {journal} {\bibinfo  {journal} {New J. Phys.}\ }\textbf {\bibinfo
  {volume} {16}},\ \bibinfo {pages} {063065} (\bibinfo {year}
  {2014})}\BibitemShut {NoStop}%
\bibitem [{\citenamefont {Fulga}\ \emph {et~al.}(2020)\citenamefont {Fulga},
  \citenamefont {Oreg}, \citenamefont {Mirlin}, \citenamefont {Stern},\ and\
  \citenamefont {Mross}}]{Fulga2020}%
  \BibitemOpen
  \bibfield  {author} {\bibinfo {author} {\bibfnamefont {I.~C.}\ \bibnamefont
  {Fulga}}, \bibinfo {author} {\bibfnamefont {Y.}~\bibnamefont {Oreg}},
  \bibinfo {author} {\bibfnamefont {A.~D.}\ \bibnamefont {Mirlin}}, \bibinfo
  {author} {\bibfnamefont {A.}~\bibnamefont {Stern}},\ and\ \bibinfo {author}
  {\bibfnamefont {D.~F.}\ \bibnamefont {Mross}},\ }\bibfield  {title} {\bibinfo
  {title} {{Temperature Enhancement of Thermal Hall Conductance
  Quantization}},\ }\href {https://doi.org/10.1103/PhysRevLett.125.236802}
  {\bibfield  {journal} {\bibinfo  {journal} {Phys. Rev. Lett.}\ }\textbf
  {\bibinfo {volume} {125}},\ \bibinfo {pages} {236802} (\bibinfo {year}
  {2020})}\BibitemShut {NoStop}%
\bibitem [{\citenamefont {Das}\ \emph {et~al.}(2023)\citenamefont {Das},
  \citenamefont {Manna},\ and\ \citenamefont {Roy}}]{SKDasPRB2023THC}%
  \BibitemOpen
  \bibfield  {author} {\bibinfo {author} {\bibfnamefont {S.~K.}\ \bibnamefont
  {Das}}, \bibinfo {author} {\bibfnamefont {S.}~\bibnamefont {Manna}},\ and\
  \bibinfo {author} {\bibfnamefont {B.}~\bibnamefont {Roy}},\ }\bibfield
  {title} {\bibinfo {title} {{Topologically distinct atomic insulators}},\
  }\href {https://doi.org/10.1103/PhysRevB.108.L041301} {\bibfield  {journal}
  {\bibinfo  {journal} {Phys. Rev. B}\ }\textbf {\bibinfo {volume} {108}},\
  \bibinfo {pages} {L041301} (\bibinfo {year} {2023})}\BibitemShut {NoStop}%
\bibitem [{\citenamefont {Das}\ and\ \citenamefont
  {Roy}(2024)}]{SKDasPRB2024THC}%
  \BibitemOpen
  \bibfield  {author} {\bibinfo {author} {\bibfnamefont {S.~K.}\ \bibnamefont
  {Das}}\ and\ \bibinfo {author} {\bibfnamefont {B.}~\bibnamefont {Roy}},\
  }\bibfield  {title} {\bibinfo {title} {Quantized thermal and spin transports
  of dirty planar topological superconductors},\ }\href
  {https://doi.org/10.1103/PhysRevB.109.195403} {\bibfield  {journal} {\bibinfo
   {journal} {Phys. Rev. B}\ }\textbf {\bibinfo {volume} {109}},\ \bibinfo
  {pages} {195403} (\bibinfo {year} {2024})}\BibitemShut {NoStop}%
\bibitem [{\citenamefont {Qi}\ \emph {et~al.}(2006)\citenamefont {Qi},
  \citenamefont {Wu},\ and\ \citenamefont {Zhang}}]{Qi2006}%
  \BibitemOpen
  \bibfield  {author} {\bibinfo {author} {\bibfnamefont {X.-L.}\ \bibnamefont
  {Qi}}, \bibinfo {author} {\bibfnamefont {Y.-S.}\ \bibnamefont {Wu}},\ and\
  \bibinfo {author} {\bibfnamefont {S.-C.}\ \bibnamefont {Zhang}},\ }\bibfield
  {title} {\bibinfo {title} {{Topological quantization of the spin Hall effect
  in two-dimensional paramagnetic semiconductors}},\ }\href
  {https://doi.org/10.1103/PhysRevB.74.085308} {\bibfield  {journal} {\bibinfo
  {journal} {Phys. Rev. B}\ }\textbf {\bibinfo {volume} {74}},\ \bibinfo
  {pages} {085308} (\bibinfo {year} {2006})}\BibitemShut {NoStop}%
\bibitem [{\citenamefont {Thouless}\ \emph {et~al.}(1982)\citenamefont
  {Thouless}, \citenamefont {Kohmoto}, \citenamefont {Nightingale},\ and\
  \citenamefont {den Nijs}}]{Thouless1982}%
  \BibitemOpen
  \bibfield  {author} {\bibinfo {author} {\bibfnamefont {D.~J.}\ \bibnamefont
  {Thouless}}, \bibinfo {author} {\bibfnamefont {M.}~\bibnamefont {Kohmoto}},
  \bibinfo {author} {\bibfnamefont {M.~P.}\ \bibnamefont {Nightingale}},\ and\
  \bibinfo {author} {\bibfnamefont {M.}~\bibnamefont {den Nijs}},\ }\bibfield
  {title} {\bibinfo {title} {{Quantized Hall Conductance in a Two-Dimensional
  Periodic Potential}},\ }\href {https://doi.org/10.1103/PhysRevLett.49.405}
  {\bibfield  {journal} {\bibinfo  {journal} {Phys. Rev. Lett.}\ }\textbf
  {\bibinfo {volume} {49}},\ \bibinfo {pages} {405} (\bibinfo {year}
  {1982})}\BibitemShut {NoStop}%
\bibitem [{SM()}]{SM}%
  \BibitemOpen
  \href@noop {} {\bibinfo  {journal} {See Supplemental Material at XXX-XXXX for
  the details of transport calculations, saturation of transport responses with
  the number of disorder realizations, calculations of electrical and thermal
  Hall conductivities, finite size effects on transport in $d+id$ paired state,
  and analytical solution of the topological zero mode}\ }\BibitemShut
  {NoStop}%
\bibitem [{\citenamefont {{S. K. Das}}\ and\ \citenamefont {{B.
  Roy}}(2024)}]{zenodoTSCSanjib}%
  \BibitemOpen
\bibfield  {journal} {  }\bibfield  {author} {\bibinfo {author} {\bibnamefont
  {{S. K. Das}}}\ and\ \bibinfo {author} {\bibnamefont {{B. Roy}}},\ }\bibfield
   {title} {\bibinfo {title} {{Quantized electrical, thermal, and spin
  transports of non-Hermitian clean and dirty two-dimensional topological
  insulators and superconductors}}\ }\href
  {https://doi.org/10.5281/zenodo.13328890} {10.5281/zenodo.13328890} (\bibinfo
  {year} {2024})\BibitemShut {NoStop}%
\bibitem [{\citenamefont {Long}\ \emph {et~al.}(2011)\citenamefont {Long},
  \citenamefont {Zhang},\ and\ \citenamefont {Sun}}]{grapheneTHC2011}%
  \BibitemOpen
  \bibfield  {author} {\bibinfo {author} {\bibfnamefont {W.}~\bibnamefont
  {Long}}, \bibinfo {author} {\bibfnamefont {H.}~\bibnamefont {Zhang}},\ and\
  \bibinfo {author} {\bibfnamefont {Q.-f.}\ \bibnamefont {Sun}},\ }\bibfield
  {title} {\bibinfo {title} {{Quantum thermal Hall effect in graphene}},\
  }\href {https://doi.org/10.1103/PhysRevB.84.075416} {\bibfield  {journal}
  {\bibinfo  {journal} {Phys. Rev. B}\ }\textbf {\bibinfo {volume} {84}},\
  \bibinfo {pages} {075416} (\bibinfo {year} {2011})}\BibitemShut {NoStop}%
\bibitem [{\citenamefont {Laughlin}(1998)}]{Laughlin1998}%
  \BibitemOpen
  \bibfield  {author} {\bibinfo {author} {\bibfnamefont {R.~B.}\ \bibnamefont
  {Laughlin}},\ }\bibfield  {title} {\bibinfo {title} {Magnetic induction of
  ${\mathit{d}}_{{\mathit{x}}^{2}\ensuremath{-}{\mathit{y}}^{2}}+{\mathit{id}}_{\mathit{xy}}$
  order in high- ${T}_{c}$ superconductors},\ }\href
  {https://doi.org/10.1103/PhysRevLett.80.5188} {\bibfield  {journal} {\bibinfo
   {journal} {Phys. Rev. Lett.}\ }\textbf {\bibinfo {volume} {80}},\ \bibinfo
  {pages} {5188} (\bibinfo {year} {1998})}\BibitemShut {NoStop}%
\bibitem [{\citenamefont {{Zhang}}\ \emph {et~al.}(2018)\citenamefont
  {{Zhang}}, \citenamefont {{Zhu}}, \citenamefont {{Zhao}}, \citenamefont
  {{Yan}},\ and\ \citenamefont {{Zhu}}}]{Discussion1}%
  \BibitemOpen
  \bibfield  {author} {\bibinfo {author} {\bibfnamefont {D.-W.}\ \bibnamefont
  {{Zhang}}}, \bibinfo {author} {\bibfnamefont {Y.-Q.}\ \bibnamefont {{Zhu}}},
  \bibinfo {author} {\bibfnamefont {Y.~X.}\ \bibnamefont {{Zhao}}}, \bibinfo
  {author} {\bibfnamefont {H.}~\bibnamefont {{Yan}}},\ and\ \bibinfo {author}
  {\bibfnamefont {S.-L.}\ \bibnamefont {{Zhu}}},\ }\bibfield  {title} {\bibinfo
  {title} {{Topological quantum matter with cold atoms}},\ }\href
  {https://doi.org/10.1080/00018732.2019.1594094} {\bibfield  {journal}
  {\bibinfo  {journal} {Adv. Phys.}\ }\textbf {\bibinfo {volume} {67}},\
  \bibinfo {pages} {253} (\bibinfo {year} {2018})}\BibitemShut {NoStop}%
\bibitem [{\citenamefont {Cooper}\ \emph {et~al.}(2019)\citenamefont {Cooper},
  \citenamefont {Dalibard},\ and\ \citenamefont {Spielman}}]{Discussion2}%
  \BibitemOpen
  \bibfield  {author} {\bibinfo {author} {\bibfnamefont {N.~R.}\ \bibnamefont
  {Cooper}}, \bibinfo {author} {\bibfnamefont {J.}~\bibnamefont {Dalibard}},\
  and\ \bibinfo {author} {\bibfnamefont {I.~B.}\ \bibnamefont {Spielman}},\
  }\bibfield  {title} {\bibinfo {title} {{Topological bands for ultracold
  atoms}},\ }\href {https://doi.org/10.1103/RevModPhys.91.015005} {\bibfield
  {journal} {\bibinfo  {journal} {Rev. Mod. Phys.}\ }\textbf {\bibinfo {volume}
  {91}},\ \bibinfo {pages} {015005} (\bibinfo {year} {2019})}\BibitemShut
  {NoStop}%
\bibitem [{\citenamefont {Asteria}\ \emph {et~al.}(2019)\citenamefont
  {Asteria}, \citenamefont {Tran}, \citenamefont {Ozawa}, \citenamefont
  {Tarnowski}, \citenamefont {Rem}, \citenamefont {Fl{\"a}schner},
  \citenamefont {Sengstock}, \citenamefont {Goldman},\ and\ \citenamefont
  {Weitenberg}}]{Discussion3}%
  \BibitemOpen
  \bibfield  {author} {\bibinfo {author} {\bibfnamefont {L.}~\bibnamefont
  {Asteria}}, \bibinfo {author} {\bibfnamefont {D.~T.}\ \bibnamefont {Tran}},
  \bibinfo {author} {\bibfnamefont {T.}~\bibnamefont {Ozawa}}, \bibinfo
  {author} {\bibfnamefont {M.}~\bibnamefont {Tarnowski}}, \bibinfo {author}
  {\bibfnamefont {B.~S.}\ \bibnamefont {Rem}}, \bibinfo {author} {\bibfnamefont
  {N.}~\bibnamefont {Fl{\"a}schner}}, \bibinfo {author} {\bibfnamefont
  {K.}~\bibnamefont {Sengstock}}, \bibinfo {author} {\bibfnamefont
  {N.}~\bibnamefont {Goldman}},\ and\ \bibinfo {author} {\bibfnamefont
  {C.}~\bibnamefont {Weitenberg}},\ }\bibfield  {title} {\bibinfo {title}
  {{Measuring quantized circular dichroism in ultracold topological matter}},\
  }\href {https://doi.org/10.1038/s41567-019-0417-8} {\bibfield  {journal}
  {\bibinfo  {journal} {Nat. Phys.}\ }\textbf {\bibinfo {volume} {15}},\
  \bibinfo {pages} {449} (\bibinfo {year} {2019})}\BibitemShut {NoStop}%
\end{thebibliography}%

\end{document}